\documentclass[aps,prd,twocolumn,twoside,superscriptaddress,floatfix, nofootinbib]{revtex4-1}
\pdfoutput=1
\normalsize
\usepackage{ifthen,amsmath,amssymb,bm,booktabs}
\usepackage{graphicx}
\usepackage{tabularx}
\usepackage[dvipsnames]{xcolor}
\usepackage{hyperref}
\usepackage{float}
\usepackage{aas_macros}
\usepackage{orcidlink}
\usepackage[normalem]{ulem}

\hypersetup{colorlinks=true,linkcolor=MidnightBlue,citecolor=MidnightBlue,filecolor=MidnightBlue,urlcolor=MidnightBlue}

\newcommand{\vl} { {\bm{\ell}} }
\newcommand{\vL} { {\bm{L}} }
\newcommand{\vx} {\bm{x}}

\newcommand{\taubias} {\hat{\tau}_{\rm bias}}
\newcommand{\beq} {\begin{equation}}
\newcommand{\eeq} {\end{equation}}
\newcommand{\bal} {\begin{aligned}}
\newcommand{\eal} {\end{aligned}}
\newcommand{\ind}{\mathrel{\perp\!\!\!\perp}} 
\newcommand{\nind}{\mathrel{\,\not\!\perp\!\!\!\perp}} 
\DeclareMathOperator{\sgn}{sgn}
\DeclareMathOperator{\var}{var}
\newcommand{\dtl} {\delta T_{\rm large}}
\newcommand{\dts} {\delta T_{\rm small}}
\newcommand{\lmax}{\ell_{\rm max}}

\begin{document}

\title{A new ``temperature inversion'' estimator to detect CMB patchy screening by large-scale structure}

\author{Theo Schutt
\orcidlink{0000-0002-7187-9628}}
\email{schutt20@stanford.edu}
\affiliation{Kavli Institute for Particle Astrophysics and Cosmology,
382 Via Pueblo Mall, Stanford, CA 94305, USA}
\affiliation{SLAC National Accelerator Laboratory, 2575 Sand Hill Road, Menlo Park, CA 94025, USA}
\affiliation{Department of Physics, Stanford University, Stanford, CA 94305, USA}

\author{Abhishek S. Maniyar
\orcidlink{0000-0002-4617-9320}}
\affiliation{Kavli Institute for Particle Astrophysics and Cosmology,
382 Via Pueblo Mall, Stanford, CA 94305, USA}
\affiliation{SLAC National Accelerator Laboratory, 2575 Sand Hill Road, Menlo Park, CA 94025, USA}
\affiliation{Department of Physics, Stanford University, Stanford, CA 94305, USA}

\author{Emmanuel Schaan
\orcidlink{0000-0002-4619-8927}}
\affiliation{Kavli Institute for Particle Astrophysics and Cosmology,
382 Via Pueblo Mall, Stanford, CA 94305, USA}
\affiliation{SLAC National Accelerator Laboratory, 2575 Sand Hill Road, Menlo Park, CA 94025, USA}

\author{William R. Coulton
\orcidlink{0000-0002-1297-3673}}
\affiliation{Center for Computational Astrophysics, Flatiron Institute, 162 5th Avenue, New York, NY 10010, USA}
\affiliation{Kavli Institute for Cosmology Cambridge, Madingley Road, Cambridge CB3 0HA, UK}

\author{Nishant Mishra
\orcidlink{0000-0002-9141-9792}}
\affiliation{Department of Astronomy, University of Michigan, Ann Arbor, MI 48109, USA}
\affiliation{Department of Physics, University of California, Berkeley, CA 94720, USA}

\begin{abstract}

Thomson scattering of cosmic microwave background (CMB) photons imprints various properties of the baryons around galaxies on the CMB.
One such imprint, called patchy screening, is a direct probe of the gas density profile around galaxies.
It usefully complements the information from the kinematic and thermal Sunyaev-Zel'dovich effects and does not require individual redshifts.
In this paper, we derive new estimators of patchy screening called the ``temperature inversion'' (TI) and ``signed'' estimators, analogous to the gradient inversion estimator of CMB lensing.
Pedagogically, we clarify the relation between these estimators and the standard patchy screening quadratic estimator (QE). The new estimators trade optimality for robustness to biases caused by the dominant CMB lensing and foreground contaminants, allowing the use of smaller angular scales. 
We perform a simulated analysis to realistically forecast the expected precision of patchy screening measurements from four CMB experiments, ACT, SPT, Simons Observatory (SO) and CMB-S4, cross-correlated with three galaxy samples from BOSS, unWISE and the simulated Rubin LSST Data Challenge 2 catalog.
Our results give further confidence in the first detection of this effect from the ACT$\times$unWISE data in the companion paper and show patchy screening will be a powerful observable for future surveys like SO, CMB-S4 and LSST.
Implementations of the patchy screening QE and the TI and signed estimators are publicly available in our \texttt{LensQuEst}\footnote{\url{https://github.com/EmmanuelSchaan/LensQuEst}} and \texttt{ThumbStack}\footnote{\url{https://github.com/EmmanuelSchaan/ThumbStack}} software packages, respectively.

\end{abstract}

\maketitle

\section{Introduction}

While the cosmic microwave background (CMB) is an extremely powerful window onto the initial conditions of our Universe, 
it also contains a wealth of information on its late-time large-scale structure (LSS).
Indeed, large-scale structure properties including gas density, thermal and bulk velocities, gravitational potential and more  leave shadows on the CMB called secondary anisotropies; see \cite{Aghanim2008Rev} for a review.
One important class of secondary anisotropies is due to Thomson scattering of the CMB photons by free electrons in the ionized plasma (gas) around galaxies and clusters.
This class includes the thermal and kinematic Sunyaev-Zel'dovich (tSZ \& kSZ) effects \cite{ZS69, SZ72, SZ80}, which are measured with high precision and have already yielded crucial insights into cosmology \cite{Salvati_2018, Bolliet_2018, Soergel:2017ahb, Amon_2022, Arico_2023.678A.109A} and the thermodynamics of gaseous halos around galaxies and clusters \citep[e.g.,][]{ Schaan:2015uaa, Schaan2021a, Amodeo2021, Schiappucci_2023}.

Another effect in this class is the patchy screening effect \citep{Ostriker_1986, Vishniac:1987wm, DvorkinSmith2009, Hernandez_Monteagudo_2010, Feng_2018}.
This effect is perhaps the simplest, most intuitive one:
a free electron along a given line of sight (LOS) scatters CMB photons out of that LOS, and replaces them with CMB photons from the average blackbody distribution, imprinting a shadow.
However, patchy screening is smaller in amplitude than the more complex tSZ and kSZ effects since they scale as
\beq
\left\{
\bal
&\delta T^{\rm tSZ} \propto \tau( v_\text{thermal}/c )^2,  &\left( v_\text{thermal}/c \right)^2 &\sim 10^{-2},\\
&\delta T^{\rm kSZ} \propto \tau(v_\text{bulk}/c),  &v_\text{bulk}/c &\sim 10^{-3},\\
&\delta T^{\rm screening} \propto \tau(\delta T_0/T_0), &\delta T_0/T_0 &\sim 10^{-4},\\
\eal
\right.
\eeq
where $\tau\sim 10^{-3}$ for galaxy group-sized halos.

These scalings also highlight the complementarity of these different effects.
Like kSZ, patchy screening is independent of the gas temperature.
It thus scales linearly with halo mass, unlike tSZ, and is thus ideally suited for constraining baryonic effects in galaxy lensing by measuring baryon density profiles around the group-sized halos that dominate the matter power spectrum \cite{Amodeo2021}.

Like tSZ, patchy screening does not require any individual redshift information to be detected.
While the kSZ effect can also be detected without redshift information \cite{Dore:2003ex, Hill:2016dta, Kusiak_2021}, 
this is done at the cost of squaring the kSZ, which destroys its linear scaling with gas density and halo mass.
Combining screening, kSZ and tSZ allows us to form a complete picture of the gas thermodynamics around halos (gas density, temperature).
Screening measurements also enables breaking of the kSZ degeneracy between galaxy gas content and galaxy velocity.
Thus, by combining them, one may extract the amplitude of large-scale peculiar velocities in the Universe to test the neutrino masses \cite{Mueller:2014dba}, the growth rate of structure \cite{Sugiyama_2017.01.057S}, general relativity \cite{Mueller:2014nsa}, and measure primordial non-Gaussianity \cite{2019PhRvD.100h3508M}.
Finally, while screening and kSZ also probe the epoch of reionization \cite{Hu:1999vq, DvorkinSmith2009, Natarajan2013.776.82N, Roy2018.05.014R, Bianchini_2023},
we will not consider this in this paper, focusing instead on the late-time signal from galaxies and clusters.

Observationally, the patchy screening effect is unique in that it changes sign depending on the sign of the local large-scale CMB temperature.
This is unlike extragalactic foregrounds like the cosmic infrared background (CIB) or the tSZ effect, whose sign are fixed at any given frequency.
This crucial feature is analogous to kSZ, whose signal changes sign based on the sign of the local LOS peculiar velocity.
Like for kSZ, it allows us to cleanly distinguish screening from these foregrounds.
This property also differentiates screening from CMB lensing, whose sign changes based on the local large-scale temperature gradient, rather than temperature value.

Leveraging this crucial feature of patchy screening, we derive new estimators dubbed ``temperature inversion'' (TI) and ``signed'' estimators, analogous to the gradient inversion (GI) estimator of CMB lensing \cite{Horowitz2019.485.3919H, Hadzhiyska2019.100b3547H}.
By making use of the sign change of the screening effect, these estimators are naturally robust to contamination from CIB, tSZ and kSZ, as well as CMB lensing.

We carefully present the relation between these new estimators and the standard quadratic estimator (QE) for patchy screening \cite{Hu:2001kj, DvorkinSmith2009, Su2011.1106.4313S}.
We show how the new estimators do not outperform the standard QE in terms of raw statistical power when using same multipole range.
However, in its simplest form, the QE is biased by foregrounds and CMB lensing.
While CMB lensing contamination to the QE can be removed via ``bias hardening,'' the situation may be more complex for extragalactic foregrounds.

While we focus on detecting screening from the CMB temperature only, these estimators generalize trivially to polarization by simply replacing temperature with Stokes Q and U.

The paper is organized as follows.
In Section \ref{sec:patchy_screening}, we review the physical origin of patchy screening.
Section \ref{sec:qe} covers the standard QE for patchy screening and its application to reconstructing both large- and small-scale optical depth $\tau$.
Section \ref{sec:ti} introduces the new TI and signed estimators and clarifies their relationship to the small-scale QE.
In Section \ref{sec:biases}, we study the biases to the QE due to lensing and extragalactic foregrounds and explain the TI estimator's robustness to these biases.
Section \ref{sec:forecasts} provides an overview of our realistic forecast simulations and analysis and presents the resulting patchy screening detection signal-to-noise ratios (SNRs) for TI and QE for a set of current and future CMB and LSS surveys.
We conclude with Section \ref{sec:conclusions}.

\section{Patchy screening of CMB temperature and polarization}
\label{sec:patchy_screening}

Consider a line of sight where the CMB temperature $T^0$ deviates from the sky averaged CMB temperature $\bar{T}^0$, by an amount $\delta T^0 = T^0 - \bar{T}^0$.
If an electron cloud with optical depth $\tau$ to Thomson scattering is present along that line of sight, two things occur.
First, a fraction $1-e^{-\tau}$ of the CMB photons with temperature $T^0$ is deflected out of the line of sight, leaving only a fraction $e^{-\tau}$ of the incident photons at temperature $T^0$.
Second, an equal fraction $1-e^{-\tau}$ of CMB photons from all other directions is scattered into the line of sight. 
Because this effect averages over all directions, as seen by the electron cloud, these photons have temperature $\bar{T}^0$.
Thus the observed temperature $T$ along the line of sight, after Thomson screening, is:
\beq
T = T^0 e^{-\tau}
+
\bar{T}^0 \left( 1 - e^{-\tau} \right).
\eeq
Considering temperature deviations with respect to the sky average CMB temperature $\bar{T}^0$, we thus obtain:
\beq
\delta T(\vx)
=
\delta T^0(\vx)
e^{-\tau(\vx)},
\eeq
where $\vx$ is the angular position on the sky.
The screening effect operates in the same way on temperature T, and Q and U polarizations.
For ease of notation, we will thus focus on temperature here, but note that all our results apply to polarization as well.

In polarization, an additional ``scattering'' effect adds to this screening effect, where the local CMB temperature quadrupole observed by the free electrons is converted to linear polarization.
This ``scattering'' occurs most noticeably at the last scattering surface and around the epoch of reionization.
It also occurs at later times due to the ionized gas around galaxies and clusters, an efect which has been called polarized Sunyaev-Ze'dovich (pSZ) effect \cite{SZ80Rev, Sazonov1999, Emritte2016}.

Going further, we decompose the optical depth field $\tau(\vx)$ into a spatial average $\bar{\tau}$, well measured by \textit{Planck} through the large-scale scattering effect ($\bar{\tau} = 0.054 \pm 0.007$ \cite{Planck2020A&A...641A...6P}), and a fluctuating part $\delta \tau$, which traces the inhomogenous projected distribution of free electrons on the sky.
Reabsorbing the mean $\bar{\tau}$ into the definition of $\delta T^0$ and assuming small fluctuations $\delta \tau (\vx) \ll 1$,
the screening effect reduces to an additive term:
\beq
\delta T^\text{screening}(\vx)
=
- \delta T^0(\vx)
\delta \tau(\vx).
\eeq

\begin{figure}[H]
\centering
\includegraphics[width= \columnwidth]{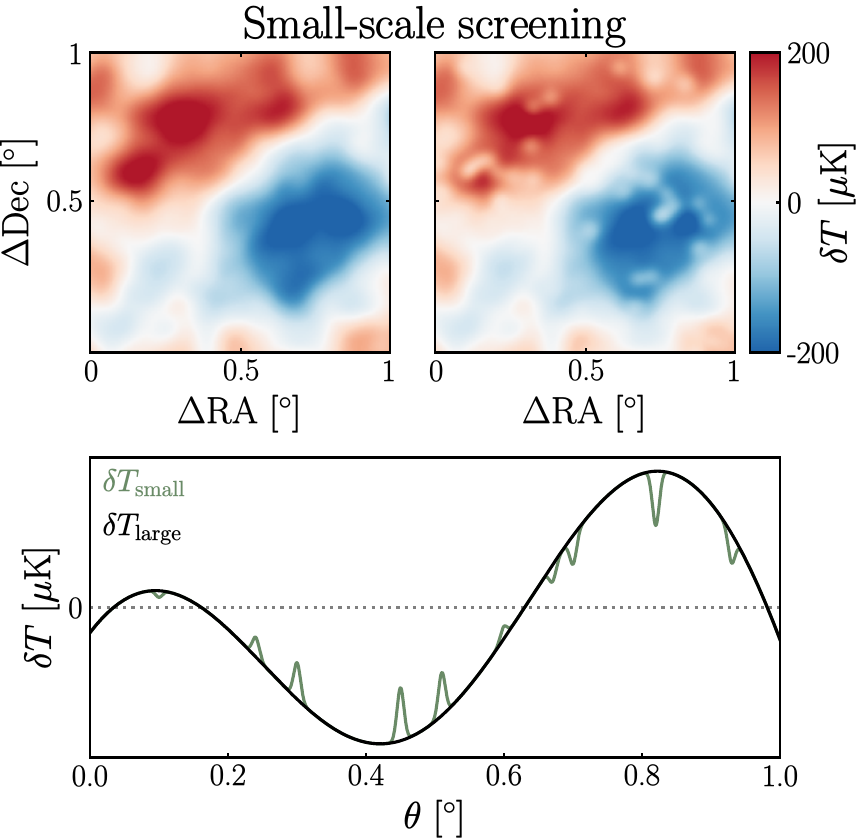}
\caption{\textbf{Top:} 1 deg$^2$ simulated map of a CMB realization unscreened (left) and screened by 5' FWHM 2D Gaussian $\tau$ profiles placed at CMASS galaxy positions (right).
The $\tau$ amplitude is greatly exaggerated for visual effect.\\
\textbf{Bottom:} A 1D schematic illustrating the small angular scale temperature deviations $\dts$ induced by the patchy screening effect (green).
Unlike any other foreground, these deviations are (anti-)correlated with the large-scale primary CMB anisotropies $\dtl$ (black). The tSZ- and CIB-induced anisotropies always have the same sign at a given frequency; lensing is correlated with $\nabla\dtl$, not $\dtl$ itself; and kSZ-induced anisotropies change sign with the gas's LOS velocity.
Thus, stacking the observed CMB temperature patches around galaxies weighted by the sign of the large-scale temperature provides a means to detect screening.
Moreover, the patchy screening effect has greater magnitude where $\dtl$ is large, motivating weighting by $\dtl$ (and not just its sign) when stacking.
}
\label{fig:illustration_screening}
\end{figure}

This effect is illustrated in Fig.~\ref{fig:illustration_screening}.
Our goal in this paper is to propose new ``temperature inversion'' and ``signed'' estimators in order to reconstruct the optical depth fluctuation field from the observed temperature map. Please note that in the discussion that follows, we use notations $\tau$ and $\delta \tau$ interchangeably, and they both represent the fluctuations on top of the mean $\bar{\tau}$ field.

\section{Standard Quadratic Estimator}
\label{sec:qe}

\subsection{Review: derivation of the standard QE}

The standard QE for $\tau$ \cite{Hu:2001kj, DvorkinSmith2009, Su2011.1106.4313S} is generally derived starting from
\beq
\delta T(\vx)
=
\delta T^0(\vx)
\left[
1 - \tau(\vx)
+\mathcal{O}(\tau^2(\vx))
\right].
\eeq
In Fourier space with a flat-sky approximation, this becomes
\beq
\delta T_\vl
\simeq
\delta T^0_\vl
-
\int_{\vl'}
\delta T^0_{\vl'}\,
\tau_{\vl - \vl '},
\eeq
where 
$\int_{\vl'} \equiv \int \frac{d^2\vl'}{(2\pi)^2}$ and $\vl$ is the 2D Fourier wavevector, which we will also refer to interchangeably as angular multipole in what follows, in analogy with curved sky analyses.
Thus screening produces non-zero off-diagonal covariances for the Fourier modes:
\beq
\langle \delta T_\vl \delta T_{\vL-\vl} \rangle_{\text{fixed }\tau}
=
f^{\tau_{L}} (\vl, \vL-\vl) \tau_\vL \, ,
\eeq
where
\beq
f^{\tau_{L}} (\vl, \vL-\vl)
=
-\left( C^0_\ell + C^0_{|\vL-\vl|} \right).
\eeq
An elementary QE for $\tau_{\vL}$ is thus simply
$- \delta T_\vl \delta T_{\vL-\vl} / \left( C^0_\ell + C^0_{|\vL-\vl|} \right)$,
with variance $C^\text{total}_\ell\ C^\text{total}_{|\vL-\vl|}\ / \left( C^0_\ell + C^0_{|\vL-\vl|} \right)$.
The standard QE for $\tau_{\vL}$ is therefore obtained from the inverse-variance linear combination of the elementary QEs,
giving
\beq
\bal
\hat{\tau}^\text{QE}_\vL
& \equiv
N_{\vL} \int_\vl F^{\tau_{L}} (\vl, \vL-\vl) \delta T_\vl \delta T_{\vL-\vl} \, \\
&= -
\frac{
\int_\vl
\frac{\left( C^0_\ell + C^0_{|\vL-\vl|} \right)}
{C^\text{total}_\ell\ C^\text{total}_{|\vL-\vl|}}\
\delta T_\vl \delta T_{\vL-\vl}
}
{
\int_\vl
\frac{\left( C^0_\ell + C^0_{|\vL-\vl|} \right)^2}
{C^\text{total}_\ell\ C^\text{total}_{|\vL-\vl|}}
}.
\label{eq:tauqe}
\eal
\eeq
where
\beq
F^{\tau_{L}} (\vl, \vL-\vl) = 
\frac{f^{\tau_{L}} (\vl, \vL-\vl)}
{C^\text{total}_\ell\ C^\text{total}_{|\vL-\vl|}} \, ,
\label{eq:weightsqe}
\eeq
and
\beq
(N_{\vL})^{-1} = \int_\vl F^{\tau_{L}} (\vl, \vL-\vl) f^{\tau_{L}} (\vl, \vL-\vl) \, ,
\eeq
where the weights $F^{\tau_{L}} (\vl, \vL-\vl)$ are determined such that the variance of Eq.~\ref{eq:tauqe}, $\mathcal{N^\tau_\vL}$, given below, is minimized. 
\beq
\mathcal{N^\tau_\vL} = 
2 (N_\vL)^2 
\int_\vl \left(F^{\tau_{L}} (\vl, \vL-\vl)\right)^2 
C^\text{total}_\ell\ C^\text{total}_{|\vL-\vl|} \, .
\label{eq:varqe}
\eeq

\subsection{Intuition: large-scale VS small-scale $\tau$ reconstruction}

The standard QE derived above (Eq.~\ref{eq:tauqe}) can be understood more simply in two limiting cases, where large-scale and small-scale Fourier modes are being reconstucted.

\begin{figure}[H]
\centering
\includegraphics[width= \columnwidth]{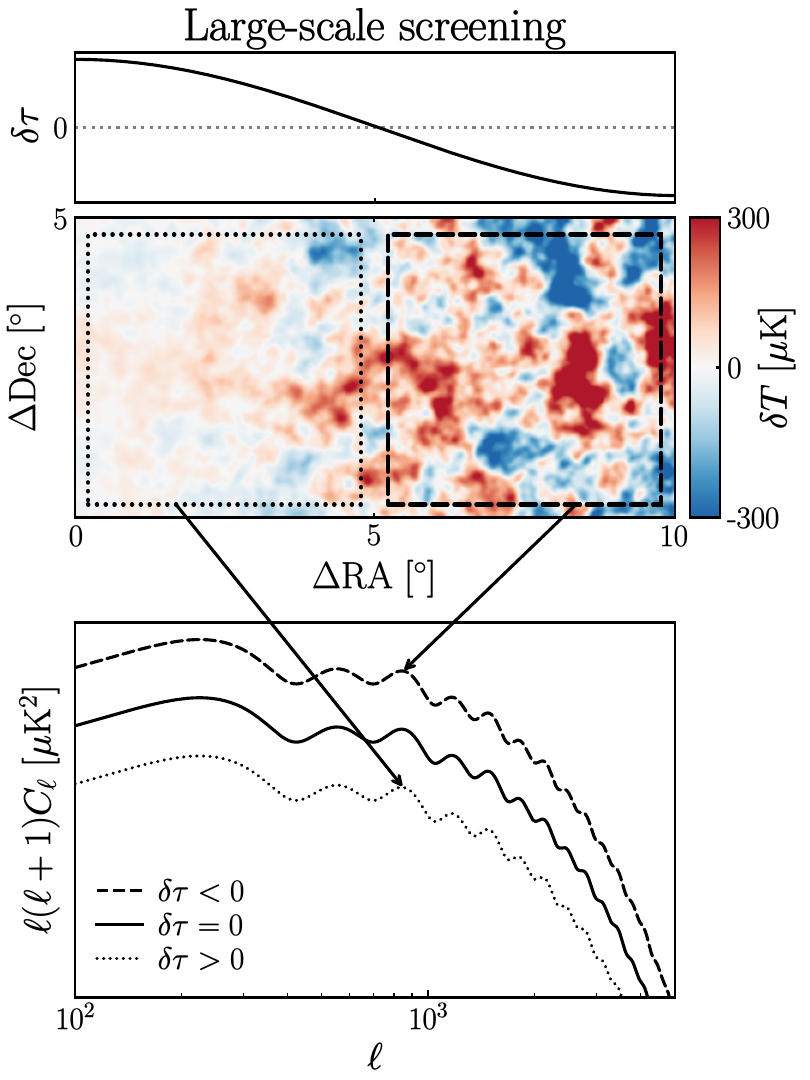}
\caption{
Schematic illustration of the effect of large-scale optical depth fluctuations on the small-scale power spectrum: the locally observed power spectrum is simply modulated in amplitude.\\
\textbf{Top:} A large-scale optical depth fluctuation $\delta \tau$ spanning 10$^{\circ}$ in RA.\\
\textbf{Middle:} A $10^{\circ} \times 5^{\circ}$ simulated temperature map of a CMB realization screened by the large-scale $\delta \tau$ fluctuation shown in the top panel. The dotted (dashed) square corresponds to the area with suppressed (enhanced) anisotropies with respect to the full-sky average optical depth $\bar{\tau}$.\\
\textbf{Bottom:} The power spectra corresponding to the two regions where $\delta\tau>0$ (dotted) and $\delta\tau<0$ (dashed). The power spectrum for the CMB screened only by $\bar{\tau}$ (i.e. $\delta\tau=0$, solid) is shown for comparison.
The screening of the CMB by a large-scale mode of the optical depth modulates the locally measured power spectrum as $C_\ell \leftarrow C_\ell e^{-2 \delta\tau}$.
The standard QE for $\delta \tau_\vL$ is looking precisely for this effect when reconstructing from temperature modes $\delta T_\vl$ with $L \ll \ell$.
}
\label{fig:illustration_large_scale_screening}
\end{figure}

In the former limit,
where large-scale $\tau_\vL$ Fourier modes are reconstructed from small-scale temperature modes $\delta T_\vl$, with $L\ll \ell$,
the standard QE is effectively looking at the local power spectrum and searching for a change of the form
\beq
C_\ell \leftarrow C_\ell e^{-2 \delta\tau},
\quad
\text{i.e.},
\quad
\frac{d\ln C_\ell}{d\tau} = -2.
\eeq
This is illustrated in Fig.~\ref{fig:illustration_large_scale_screening}.

In the latter limit,
where we reconstruct the optical depth field $\tau_\vL$ on small scales, i.e., $L\gg \ell$,
we can approximate the power spectra as smoothly-varying functions:
$C^0_{|\vL-\vl|} \simeq C^0_L$ and
$C^\text{total}_{|\vL-\vl|} \simeq C^\text{total}_L$ to lowest order.
Furthermore, the CMB power spectrum is very red, such that
$C^0_\ell \gg C^0_L$.
In this limit, the standard QE takes the suggestive form
\beq
\hat{\tau}^\text{QE}_\vL
\simeq
-
\frac{
\int_\vl
\left( \frac{C^0_\ell}{C^\text{total}_\ell} \delta T_\vl \right)
\delta T_{\vL-\vl}
}
{
\int_\vl
\frac{\left(C^0_\ell\right)^2}
{C^\text{total}_\ell}
},
\eeq
or in configuration space:
\beq
1-\hat{\tau}^\text{QE}(\vx)
\simeq
\delta T(\vx) \
\frac{\delta T^\text{large-WF}(\vx)}
{\langle \left(\delta T^\text{large-WF}\right)^2 \rangle}.
\eeq
In this limit, the standard QE is thus simply multiplying the local small-scale temperature $\delta T(\vx)$
by the Wiener-filtered large-scale temperature $\delta T^\text{large-WF}$, and normalizing it by the expected variance of the Wiener filter.
The significance of this expression will become clear in Sec.~\ref{sec:ti}, where we connect it to the new TI and signed estimators.

\section{``Temperature-inversion" estimator \& approximate small-scale QE}
\label{sec:ti}

\subsection{New ``temperature-inversion'' estimator}

The optical depth fluctuation field possesses power on very small scales since it traces galaxy halos, whereas the primary CMB temperature has almost none due to Silk damping.
Denoting small and large scales with subscripts $S$ and $L$ respectively, the screening effect on small scales is thus almost exclusively sourced by the following configurations:
\beq
\delta T^\text{screening}_S(\vx)
=
- \delta T^0_L(\vx)
\delta \tau_S(\vx).
\eeq
This separation of scales is crucial:
thanks to it, both the unscreened and screened temperature fluctuations can be measured from the same temperature map \cite{Natarajan2013.776.82N}.

Indeed, we can low- and high-pass filter the observed temperature map to estimate the large and small scale temperature fluctuations $\delta \hat{T}_L(\vx)$ and $\hat{\delta T_S}(\vx)$.
A natural, if somewhat singular, estimator for $\tau$ is then obtained by simply taking the ratio:
\beq
\hat{\tau}^\text{TI}(\vx)
\equiv
-
\frac{\delta \hat{T}_S(\vx)}{ \delta \hat{T}_L(\vx)} .
\label{eq:elementary_ti_singular}
\eeq
We call this the ``temperature inversion'' (TI) estimator in analogy with the gradient inversion (GI) estimator of CMB lensing \cite{Horowitz2019.485.3919H, Hadzhiyska2019.100b3547H}.
A natural choice of low-pass filter in this case would be a Wiener filter, the linear estimator for $\delta T^0_L(\vx)$ that has the lowest mean-squared error.
A natural choice for the high-pass filter would be an inverse-variance weighted average of the small scales, the unbiased linear estimator of $\delta \hat{T}_S(\vx)$ with the lowest variance.
In general, we can define linear filters $W_L$ and $W_S$ such that
\beq
\left\{
\bal
&\hat{\delta T_L}(\vx) = \int  d\vx' W_L (\vx - \vx') {\delta T}(\vx),\\
&\hat{\delta T_S}(\vx) = \int d\vx' W_S (\vx - \vx') {\delta T}(\vx).\\
\eal
\right.
\eeq
In practice, we apply low- and high-pass filters (see Sec. \ref{sec:ti_qe_sims}) to the temperature map to isolate $\delta T_L$ and $\delta T_S$, respectively.
If the shape of the $\tau$ profile for the objects of interest is known (e.g. from kSZ, which has higher SNR), we can use a matched filter with this profile to estimate the amplitude of the signal from the data.
However, our filters, with variable size, allow us to actually measure the scale dependence of the electron density profile. 
When jointly analysed, including their covariance matrix, they are as statistically optimal as the matched filter.

In App.~\ref{app:optimal_tlarge_estimator}, we further derive the filter $W_L$ which minimizes the mean-squared error in the presence of foreground contamination with known profiles (a combination of matched filter and Wiener filter), but do not implement it in what follows.

\subsection{Comparison to the approximate small-scale QE}

In the limit where we are reconstructing small-scale optical depth fluctuations, we have shown that the QE takes the approximate form
\beq
1-\hat{\tau}^\text{QE}(\vx)
\simeq
\delta T(\vx) \
\frac{\delta T^\text{large-WF}(\vx)}
{\left<\left(\delta T^\text{large-WF}\right)^2\right>},
\label{eq:QErealspace}
\eeq
where $\delta T^\text{large-WF}$ is the Wiener filter estimator for the unscreened CMB, given the map noise. This can be contrasted with our TI estimator of the form
\beq
1-\hat{\tau}^\text{TI}(\vx)
\simeq
\frac{\delta T(\vx)}
{\delta T^\text{large-WF}(\vx)}.
\label{eq:TIrealspace}
\eeq
In other words, in the small-scale limit, the two estimators are related via
\beq
1-\hat{\tau}^\text{QE}(\vx)
=
\left[ 1-\hat{\tau}^\text{TI}(\vx) \right]
\frac{\left(\delta T^\text{large-WF}(\vx)\right)^2}{\left<\left(\delta T^\text{large-WF}\right)^2\right>}
.
\label{eq:QETIrelation_realspace}
\eeq
Thus, even if the large-scale temperature fluctuation $\delta T^\text{large-WF}$ is measured with infinite precision, the QE suffers from an irreducible variance due to the ratio between $(\delta T^\text{large-WF}(\vx))^2$ and its mean.
This is completely analogous to the GI estimator of CMB lensing.
In App.~\ref{app:variance_comparison_ti_qe}, we expand on this comparison, showing that the SNR is higher for TI than the QE when the large-scale temperature leg is measured at high SNR (already true for \textit{Planck} data) and the small-scale temperature leg is dominated by the screening effect.
This latter condition is likely not satisfied; even in the absence of detector noise, and assuming that all non-blackbody foregrounds have been subtracted, kSZ and CMB lensing will likely dominate the noise unless they can be largely subtracted.

If TI does not outperform the QE in terms of SNR when considering the same set of temperature Fourier modes, why consider it further?
As we show in Secs. \ref{sec:ti_lens} and \ref{sec:ti_fg}, TI is more robust to extragalactic foreground biases and naturally robust to CMB lensing bias.
This means that TI can use more temperature Fourier modes (higher $\ell_\text{max}^{T}$) and thus potentially recover a higher total SNR while still being as or more robust to biases.

\subsection{Stacked TI estimator for cross-correlation \& comparison with small-scale QE}
\label{sec:ti_stack}

For a single galaxy, the na\"ive TI estimator (Eq.~\eqref{eq:elementary_ti_singular}) is clearly singular for lines of sight where the unscreened temperature fluctuation is zero.
This issue is automatically addressed when stacking around a set of galaxies, as we now show.
The large-scale temperature $\hat{\delta T_L}(\vx)$ can be measured to high precision from the temperature map, such that we ignore the noise on $\hat{\delta T_L}(\vx)$ for now.
On the other hand, the small-scale temperature $\hat{\delta T_S}(\vx)$ is typically affected by a non-negligible noise $n_S$ (with variance $\sigma_{n_S}^2$), such that
\beq
\hat{\tau}^\text{TI}(\vx)
=
\tau(\vx)
+
\frac{n_S(\vx)}{\hat{\delta T_L}(\vx)}.
\eeq
In this case, the estimator is unbiased, and its noise is smallest along lines of sight with the highest large-scale temperature, as expected intuitively.

To extract the mean optical depth profile of a sample of tracers (galaxies, galaxy groups or clusters), we look for the unbiased linear combination of the $\hat{\tau}^\text{TI}$ around each cluster that has the smallest noise variance.
The answer is the inverse-variance weighted average:
\beq
\hat{\tau}^\text{TI stack}
\equiv
\frac{
\sum_i \hat{\tau}^\text{TI}_i \times \hat{\delta T_L}_i^2 / \sigma_{n_S}^2
}
{
\sum_i \hat{\delta T_L}_i^2 / \sigma_{n_S}^2
}
,
\eeq
which can be rewritten as
\beq
\hat{\tau}^\text{TI stack}
=
-\frac{
\sum_i
\hat{\delta T}_{S i}\hat{\delta T_{Li}} / \sigma_{n_{S}}^2
}
{
\sum_j \hat{\delta T_L}_j^2 / \sigma_{n_{S}}^2
}
.
\label{eq:def_stacked_TI}
\eeq
Conveniently, the inverse-variance weighting regularizes the singularity of the na\"ive estimator, such that lines of sight where the unscreened CMB temperature $\hat{\delta T^0}$ is estimated to vanish are given zero weight in the sum.

Although the numerator of Eq.~\eqref{eq:def_stacked_TI} is now quadratic in the temperature map, the estimator as a whole is not  since the denominator also involves a quadratic combination of the data.
As derived in App.~\ref{app:stacked_TI_vs_QE}, the stacked QE takes a very similar form in the limit of small-scale screening:
\beq
\hat{\tau}^\text{QE stack}
=
-\frac{
\sum_i
\hat{\delta T}_{S i}\hat{\delta T_{Li}} / \sigma_{n_{S}}^2
}
{
\sum_j \sigma_{T_L}^2 / \sigma_{n_{S}}^2
}
.
\label{eq:def_stacked_QE}
\eeq
The numerator is identical to that of the stacked TI, and the denominator only differs by replacing $\hat{\delta T_L}^2_j$ with $\sigma_{T_L}^2$.
Thus, the ratio of the stacked QE to the stacked TI is simply
\beq
\frac{\hat{\tau}^\text{QE stack}}{\hat{\tau}^\text{TI stack}}
=
\frac{\sum_j \hat{\delta T_L}_j^2 / \sigma_{n_{\S}}^2}
{\sum_j \sigma_{T_L}^2 / \sigma_{n_{S}}^2}.
\eeq
This ratio has mean 1, but its variance is non-zero, equal to $2/N_\text{galaxies}$.
Thus, in the limit of zero temperature noise, the stacked QE retains an irreducible variance $2 \sigma_\tau^2/N_\text{galaxies}$, whereas the stacked TI yields a noiseless reconstruction of the true $\tau$.
While this variance vanishes as $\propto 1/N_\text{galaxies}$ for large galaxy samples, the other sources of variance (e.g., from detector noise) also share the same scaling, such that this additional noise remains relevant for all sample sizes.

\subsection{``Signed estimator": suboptimal to TI, but potentially even more robust to foregrounds}
\label{sec:signed_estimator}

The key intuition behind the derivation of the stacked TI estimator (Eq.~\eqref{eq:def_stacked_TI}) is shown in Fig.~\ref{fig:illustration_screening}:
unlike all other imprints of galaxies on the CMB, the screening signal changes sign and amplitude depending on those of the local large-scale temperature $\delta T_L$.
Discarding the amplitude information and weighting only by the sign of $\delta T_L$ leads to a different screening estimator.
We name it the ``signed'' estimator:
\beq
\hat{\tau}^{\rm sgn} \equiv \frac{\sum_i\sgn(\delta T_{L,i})\delta T_{S,i}}{\sum_i|\delta T_{L,i}|},
\label{eq:def_stacked_signed}
\eeq
where $\delta T_{L,i}$ (or $\delta T_{S,i}$) is the large-scale (or small-scale) temperature fluctuation at the position of galaxy $i$.

While the TI and signed estimators are both unbiased estimators of $\tau$, their variances differ, such that
\beq
\frac{\text{var}\ \hat{\tau}^{\rm sgn}}
{\text{var}\ \hat{\tau}^{\rm TI}}
=
\frac{\langle \left(\delta T_{L, i}\right)^2 \rangle}
{\langle | \delta T_{L, i} | \rangle^2}
=
\frac{\pi}{2}
,
\eeq
for the Gaussian distributed large-scale temperatures $\delta T_{L}$.
Thus, the signed estimator has lower SNR than TI by a factor 
$\sqrt{\pi/2} = 1.25$.

On the other hand, this signed estimator enables two interesting avenues for controlling potential foregrounds, which may be worth the cost in SNR:
\begin{itemize}
\item \textbf{Sign balancing.}
The number of galaxies coinciding with either positive or negative CMB patches will follow a binomial distribution, and thus will not perfectly cancel for a finite galaxy sample. We can discard a very small number of galaxies on CMB patches with the excess sign to ensure that
$\sum_i\sgn(\delta T_{L,i}) = 0$.
Performing this ``sign balancing'' step exactly nulls any part of the foreground emission (e.g., CIB, tSZ) that is constant across galaxies.
This does not lead to any bias in the estimated screening since this selection depends on the large-scale temperature only, statistically independent from the screening signal.
Furthermore, this typically only requires discarding a fraction $\sim 1/N_\text{gal}$ of the galaxy sample, completely negligible for the large galaxy catalogs of interest here.
This is described in more detail in App.~\ref{app:sign_balancing_foreground_cancellation}.
\item \textbf{Sign thresholding.}
After masking bright point sources and clusters from the temperature map, any residual foreground emission from our galaxies will be small ($\sim \mu$K) compared to the typical large-scale temperature fluctuations ($\sim 100 \mu$K for the CMB).
Thus, we may select for the stack only the objects for which $|\delta T_{L,i}|$ is larger than a few times the typical foreground signal (e.g., ($\sim 10 \mu$K)).
This ensures that residual foregrounds are not large enough to affect the sign weighting used in the signed estimator.
This is in contrast with TI, where residual foregrounds in $\delta T_{L,i}$, however small, remain present. This is described in more detail in App. \ref{app:foreground_bias_ti_sign}.
\end{itemize}
Including tests on simulations for biases from extragalactic sources, this estimator is presented in detail in \cite{Coulton_2024}, where we test for foreground biases and validate these arguments by applying the signed estimator to realistic, non-Gaussian sky simulations.

\section{Biases to QE \& TI from CMB lensing and extragalactic foregrounds}
\label{sec:biases}

\subsection{CMB lensing bias to QE: lens hardening needed}
\label{sec:qe_lens}

From Eq.~\eqref{eq:tauqe}, we can see that the QE for $\tau$ is of the form 
\beq
\label{eq:tauqe_brief}
\hat{\tau}^\text{QE}_\vL
\equiv
- N_L 
\int_\vl
\frac{\left( C^0_\ell + C^0_{|\vL-\vl|} \right)}
{C^\text{total}_\ell\ C^\text{total}_{|\vL-\vl|}}\
\delta T_\vl \delta T_{\vL-\vl} \, .
\eeq
This has a similar form as that of the QE for the lensing potential: $\hat{\phi}^\text{QE}_\vL \propto 
\int_\vl
F(\vl, \vL-\vl) \delta T_\vl \delta T_{\vL-\vl}$,
where $F(\vl, \vL-\vl)$ are optimized weights for the pair of multipoles $\delta T_\vl$ and $\delta T_{\vL-\vl}$, such that the variance of the estimator is minimized. 
Thus, Eq.~\eqref{eq:tauqe} can be interpreted as an estimator for the CMB lensing potential reconstruction with non-optimized weights, and as such, has a non-zero response to lensing. 
We can rewrite Eq.~\eqref{eq:tauqe_brief} as
\beq
\langle \hat{\tau}^\text{QE}_\vL \rangle 
= 
{\tau}^\text{QE}_\vL 
+ N_L 
\underbrace{\left(\int_\vl F^\tau(\vl, \vL-\vl) f^\phi_{TT}(\vl, \vL-\vl)\right)}
_\text{Response of QE to lensing}
\phi(\vL) \, ,
\label{eq:response_QE_lensing}
\eeq
where we have used $\langle \delta T_\vl \delta T_{\vL-\vl} \rangle = f^\phi_{TT}(\vl, \vL-\vl) \phi(\vL)$ with $f^\phi_{TT}(\vl, \vL-\vl) = C^0_\ell (\vL \cdot \vl) + C^0_{|\vL-\vl|} (\vL \cdot (\vL - \vl))$, and $F^\tau(\vl, \vL-\vl)$ is given by Eq.~\eqref{eq:weightsqe}. 
In practice, the lensing bias in Eq.~\eqref{eq:response_QE_lensing} is non-zero and can be dominant even in cross-correlations with other tracers, as shown in Tab.~\ref{tab:foreground_bias_qe} and Fig.~\ref{fig:QEbiasplot} (see also Fig. 1 of \cite{Namikawa_2021}). 

This large lensing bias to the QE is typically avoided via ``lens hardening'' \cite{Namikawa_2013}.
The idea behind the lens-hardened QE is simple. 
Just as we can construct a $\hat{\tau}$ QE, we can also construct a lensing convergence $\hat{\kappa}$ QE, which can then be used to subtract off the lensing bias in Eq.~\eqref{eq:response_QE_lensing}. 
Therefore, the lens-hardened $\tau$ QE minimizes contamination from lensing. 
However, this comes at an increased reconstruction noise cost when compared with the standard QE variance $\mathcal{N^{\tau}_\vL}$ given by Eq.~\ref{eq:varqe}. 
For a \textit{Planck}-like experiment, this noise cost is $\sim 0-40\%$ depending on scales \cite{Namikawa_2021}, whereas the reconstruction noise from temperature-only (no polarization) QE increases by a factor of $\sim 2-3$ across all $\ell$ values for a CMB-S4-like experiment \cite{Roy_2023}.

Intuitively, this lensing bias to QE can be understood from Fig.~\ref{fig:illustration_large_scale_screening} when reconstructing large-scale optical depth fluctuations $\tau_\vL$ from small-scale temperature modes $T_\vl$ with $L \ll \ell$.
Both lensing and screening cause variations in the local power spectrum from patch to patch.
While these variations are different, they have non-zero overlap.

The opposite regime, where we reconstruct small-scale optical depth fluctuations $\tau_\vL$ from temperature modes $T_\vl$ and $T_{\vL-\vl}$ with $L \gg \ell$, is different. In this limit,
the lensing term in Eq.~\eqref{eq:response_QE_lensing} vanishes.
Thus, the small-scale part of the QE is unbiased with respect to lensing.
This part of the QE is very similar to the TI estimator, only with a slightly different normalization (see Eqs.~\eqref{eq:QErealspace}-\eqref{eq:QETIrelation_realspace}), such that both are unbiased by lensing.
We explore the robustness of TI to CMB lensing in more detail in the next subsection.

\subsection{TI is robust to CMB lensing}
\label{sec:ti_lens}

The lensed temperature map can be written, to first order, as
\beq
\delta \Tilde{T}(\vx) \approx 
\delta T(\vx) + \nabla \phi(\vx) \nabla \delta T(\vx),
\eeq
where $\tilde{T}$ is the lensed temperature field and $\phi$ is the lensing potential.
Thus, lensing by a galaxy of interest produces a local small-scale dipole in the CMB map, whose direction and amplitude matches those of the local large-scale unlensed CMB gradient.
The spatial symmetry of this dipole effect means that lensing is nulled by our azimuthally symmetric small-scale temperature filter, typically a disk or ring aperture around the galaxy, which only extracts the local small-scale monopole.
Thus, to first order, lensing from the galaxies in our sample does not bias the TI or signed estimators and does not even contribute noise.

The situation is different for the lensing caused by objects not aligned with the aperture filter.
For a single galaxy, nearby objects (including other galaxies in the sample) can contribute a lensing dipole to the aperture photometry.
Since these small-scale dipoles are no longer aligned with our small-scale apertures (rings or disks), lensing does enter the TI and signed estimators.
This lensing is not a bias, however, as it is correlated with $\nabla \delta T_L$, a random variable uncorrelated with $\delta T_L$ or $\delta T_S$.
Since there is an equal number of galaxies on either side of the gradient on average, lensing from objects not aligned with the filter only contributes noise to the TI and signed estimators.
In fact, on small scales $\ell \gtrsim 5000$, the CMB power spectrum is dominated by lensing, meaning that lensing may be the dominant source of noise (other than kSZ, other foregrounds and detector noise potentially present in the map).

In summary, TI and signed estimators null the lowest order bias due to lensing $\sim \mathcal{O} \left( \phi g \right)$.
At higher order in lensing ($\mathcal{O} \left( \phi^2 \right)$), a bias to the TI and signed estimators may be present if
\begin{itemize}
\item lensing is non-negligible in the large-scale temperature map as well;
\item this lensing is not due to the galaxy we are centered on, such that the corresponding dipole is not nulled by our small-scale aperture filter;
\item the lensing is still caused by objects correlated with the galaxy we are centered on, e.g., two-halo term or other correlated galaxies in the sample. 
\end{itemize}
We leave the exploration of these higher order bias terms in simulations to future work.

\subsection{Extragalactic foregrounds bias QE}

\begin{figure}
\centering
\includegraphics[width= \columnwidth]{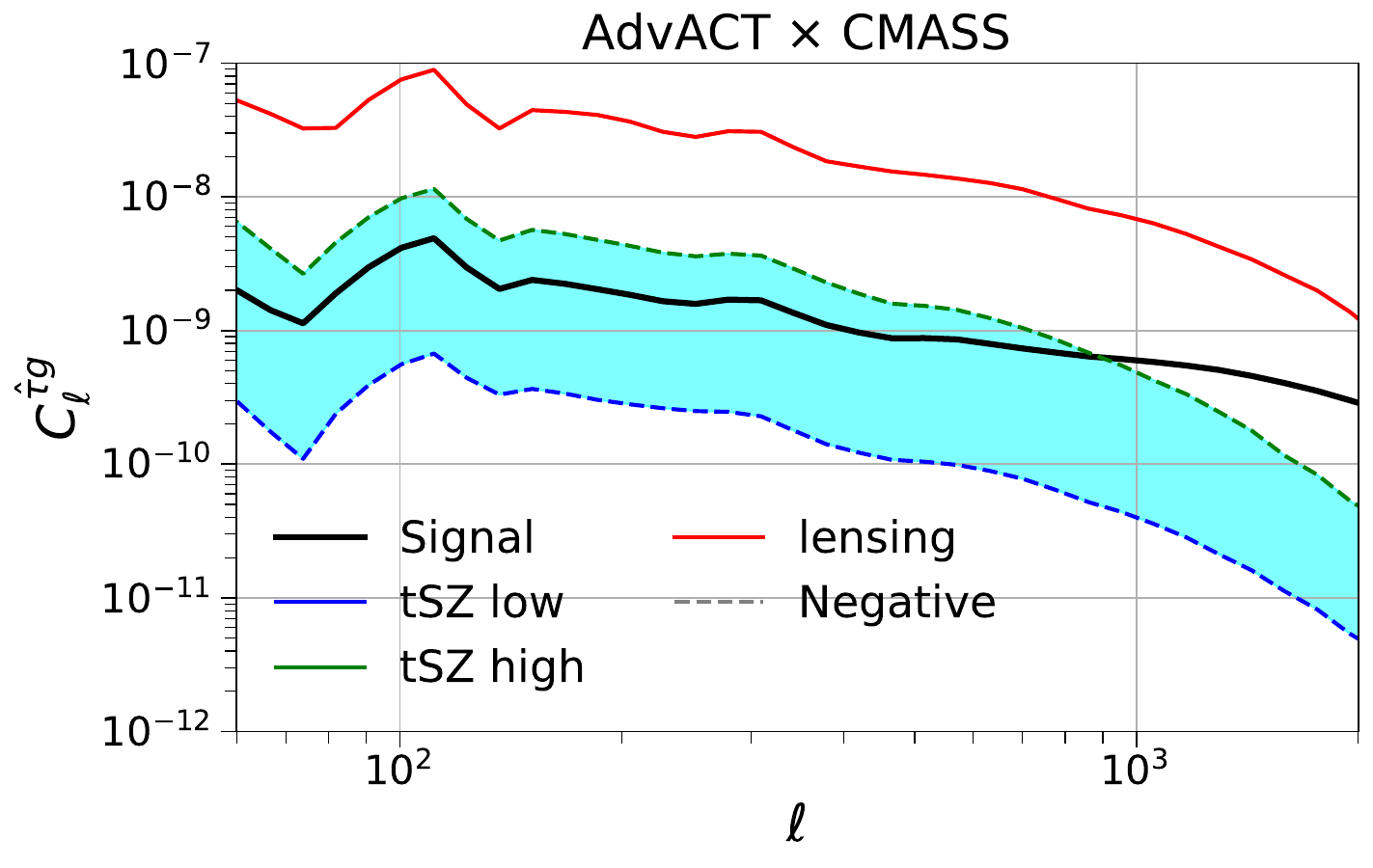}
\caption{
The expected cross-correlation signal between BOSS CMASS galaxies and the screening QE from Advanced ACT-like temperature (black) is compared to the CMB lensing bias (red) and the tSZ foreground bias (cyan area) for a range of tSZ models.
The importance of these biases highlight the need for lens hardening and foreground hardening for the $\tau$ QE.
The corresponding biases to the overall amplitude of the signal are shown in Tab.~\ref{tab:foreground_bias_qe}. The simulations used to create these correlated lensing and tSZ maps are described in detail in App.~\ref{app:fgsim}.
}
\label{fig:QEbiasplot}
\end{figure}

Similar to CMB lensing, foregrounds like tSZ, kSZ and CIB cause mode coupling of temperature anisotropies. 
As a result, they bias the QE for screening. 
Again, similar to CMB lensing, this bias also affects cross-correlation of the reconstructed $\tau$ field with other tracers. 
Fig.~\ref{fig:QEbiasplot} shows the expected cross-correlation of the $\tau$ field reconstructed from a QE using an Advanced ACT-like experiment with CMASS galaxies. Here, ``tSZ high'' and ``tSZ low'' are two tSZ maps we simulate which give an upper and lower bound respectively on this bias (see App.~\ref{app:fgsim} for details). We can see that the bias from CMB lensing is higher than the signal itself. The bias from tSZ emission is not negligible either. While we only discuss the foreground bias due to CMB lensing and tSZ here, foregrounds like CIB could also have non-negligible bias to $\tau$ QE. Such foreground biases have been extensively studied and shown to be important in the context of CMB lensing QE \cite{vanEngelen_2014, Osborne_2014}. Similar conclusions hold for the $\tau$ QE as well due to the similarity between the two estimators. It should also be noted that these biases are likely bigger for lower noise CMB experiments as they will have increased weights from small scales where the foregrounds are dominant.

In order to quantify the bias from foregrounds like CMB lensing and tSZ to the $\tau$ QE in cross-correlation with galaxies, we define an amplitude parameter $A$ for a given power spectrum measurement $\hat{C}_L$ such that
\beq
\hat{C}_L = A\  C^{\rm theory}_L + n_L \, ,
\eeq
where $C^{\rm theory}_L$ is the theory power spectrum and $n_L$ is noise. 
The minimum-variance estimator for $A$, linear in the measured power spectrum, is then
\beq
\hat{A}
= \frac{\sum_L \hat{C}_L C^{\rm theory}_L / {\sigma_{L}}^2}{\sum_L ({C^{\rm theory}_L})^2 / {\sigma_{L}}^2} \, .
\eeq
The nominal value of $\hat{A}$ is 1; however, due to biases from foregrounds, the value differs from 1. 
Thus, bias from a foreground $X$ to cross-correlation of the QE reconstructed $\tau$ field with galaxies $\delta_g$ ($C_L^{\hat{\tau} \times \delta_g}$) can be calculated in the continuum limit as
\beq
{\rm bias}(C^{\hat{\tau} \times \delta_g}) = \frac{\int \frac{d^2 {\bold L}}{(2\pi)^2} \frac{{\rm bias}(C_L^{X \times \delta_g})}{C_L^{\hat{\tau} \times \delta_g}} \frac{(C_L^{\hat{\tau} \times \delta_g})^2}{\sigma_L^2}}{{\int \frac{d^2 {\bold L}}{(2\pi)^2} \frac{(C_L^{\hat{\tau} \times \delta_g})^2}{\sigma_L^2}}} \, ,
\eeq
where ${\rm bias}(C_L^{X \times \delta_g})$ is the bias introduced by the cross-correlation between the galaxies
and the foreground $X$ passed through a $\tau$ QE.
Tab.~\ref{tab:foreground_bias_qe} shows that these foreground biases can be larger than the screening signal and highly significant.
Some form of foreground mitigation is therefore likely required for the QE.
This may include multi-frequency component separation, masking, template deprojection, using shear only estimators, and bias hardening with respect to point sources or specific profiles. Such approaches and their challenges have been explored in the context of CMB lensing and $\tau$ QE  \cite{Namikawa_2013, Namikawa_2021, MacCrann_2023, Schaan_2019, Sailer_2020} and similar results hold here too. 
We do not explore these further in this paper.
\begin{table}[H]
    \centering
    \begin{tabular}{@{}llllll@{}}
    \toprule
      \begin{tabular}{l}
      Galaxy\\ sample
      \end{tabular} &
      \begin{tabular}{l}
      Fore-\\ground
      \end{tabular} &
      \begin{tabular}{l}
        SPT
      \end{tabular} &
      \begin{tabular}{l}
        ACT
      \end{tabular} &
      \begin{tabular}{l}
        SO
      \end{tabular} &
      \begin{tabular}{l}
        CMB-S4
      \end{tabular}
      \\
      \midrule
      \begin{tabular}{l}
        CMASS
      \end{tabular} &
      \begin{tabular}{l}
        $\kappa$ \\
        tSZ low\\
        tSZ high\\
      \end{tabular} &
      \begin{tabular}{l}
        410.9\% \\
        -2.7\% \\
        -39.7\% \\
      \end{tabular} &
      \begin{tabular}{l}
        346.9\% \\
        -2.5\% \\
        -28.4\% \\
      \end{tabular} &
      \begin{tabular}{l}
        394\% \\
        -2.7\% \\
        -36.5\% \\
      \end{tabular} &
      \begin{tabular}{l}
        433.5\% \\
        -2.8\% \\
        -44.2\% \\
      \end{tabular} \\
    \bottomrule
    \end{tabular}
    \caption{
    The amplitude of the cross-correlation of galaxies with the screening QE suffers large fractional biases from CMB lensing and tSZ, using $30<\ell<3000$.
    App. \ref{app:tszsims} describes the ``low'' and ``high'' tSZ modeling in detail.
    }
    \label{tab:foreground_bias_qe}
\end{table}

\subsection{TI is robust to extragalactic foregrounds}
\label{sec:ti_fg}

The TI and signed estimators are automatically robust to extragalactic foreground biases, and this can be further enhanced by sign balancing and sign thresholding for the signed estimator.
We summarize the reasons here and present more detailed derivations in App.~\ref{app:sign_balancing_foreground_cancellation} and~\ref{app:foreground_bias_ti_sign}.

Unlike the patchy screening effect, foregrounds like the CIB, tSZ and kSZ do not change sign depending on the sign of the local large-scale temperature $\delta T_L$.
In the absence of foreground contamination to the large-scale leg,
the weights in the TI and signed estimators, which change sign and have zero mean value, automatically average out these foregrounds.
For a finite galaxy sample size, this foreground reduction may leave a residual bias
$\sim f_S / \sqrt{N_\text{patch}}$, where $f_S$ is the typical small scale foreground emission, and 
$N_\text{patch}$
is the number of patches with independent large-scale CMB temperature $\delta T_L$.
Since the CMB is coherent on degree scales, $N_\text{patch}$ is roughly equal to the area of the CMB map in square degrees.

For the signed estimator, this residual can be further reduced by sign balancing the weights, as described in Sec.~\ref{sec:signed_estimator}.
If the foreground signal is identical for all galaxies in the sample, sign balancing exactly nulls the foreground bias.
While foreground emissions actually vary across galaxies, sign balancing is still effective in realistic scenarios, as shown in App.~\ref{app:sign_balancing_foreground_cancellation}

In practice, foreground contamination may be present both in the large and small-scale temperature legs. 
In this case, the TI and signed estimators may suffer a bispectrum bias 
$\sim f_L f_S g_S$, 
where $f$ is the foreground and $g$ the galaxy map.
However, for the signed estimator, this bias can be nulled exactly with the sign thresholding procedure presented  in Sec.~\ref{sec:signed_estimator}.
This procedure ensures that absolutely no foreground signal leaks into $\text{sgn}\left(\delta T_L\right)$, thus removing any bispectrum foreground bias.

\section{SNR forecasts}
\label{sec:forecasts}

We produce SNR forecasts for the TI and $\tau$ QE estimators using realistic simulations for four CMB experiments, South Pole Telescope (SPT) \cite{Benson_2014, 2018SPIE10708E..03B, Sobrin_2022}, Atacama Cosmology Telescope (ACT) \cite{2016JLTP..184..772H}, Simons Observatory (SO) \cite{SOScience} and CMB-S4 \cite{2019arXiv190704473A}, cross-correlated with three galaxy survey samples, the Baryonic Oscillation Spectroscopic Survey (BOSS) CMASS sample \cite{Dawson_2013, Reid_2016}, the Wide-field Infrared Survey Explorer (WISE) unWISE ``blue'' and ``green'' samples \cite{Wright_2010, Meisner_2018}, and the simulated Legacy Survey of Space and Time (LSST) Data Challenge 2 (DC2) extragalactic sample \cite{DC2_2021}.
Details of the sky fraction, galaxy number density and number of galaxies overlapping the CMB survey footprints are listed in Tab.~\ref{tab:ngal}.

\subsection{Simulated analysis for TI and QE}
\label{sec:ti_qe_sims}
Here we give an overview of the methods used to generate the simulations for the TI and QE detection forecast.
See App.~\ref{app:tisims} for further details.
We create realistic simulated $\tau$ signal maps by placing identical, isotropic Gaussian $\tau$ profiles at CMASS galaxy positions as described in App.~\ref{app:tisims_signal}.
The parameters of the $\tau$ profiles are matched to the best-fit values from the ACT$\times$unWISE first measurement \cite{Coulton_2024}.

\begin{figure}[h]
\centering
\includegraphics[width= \columnwidth]{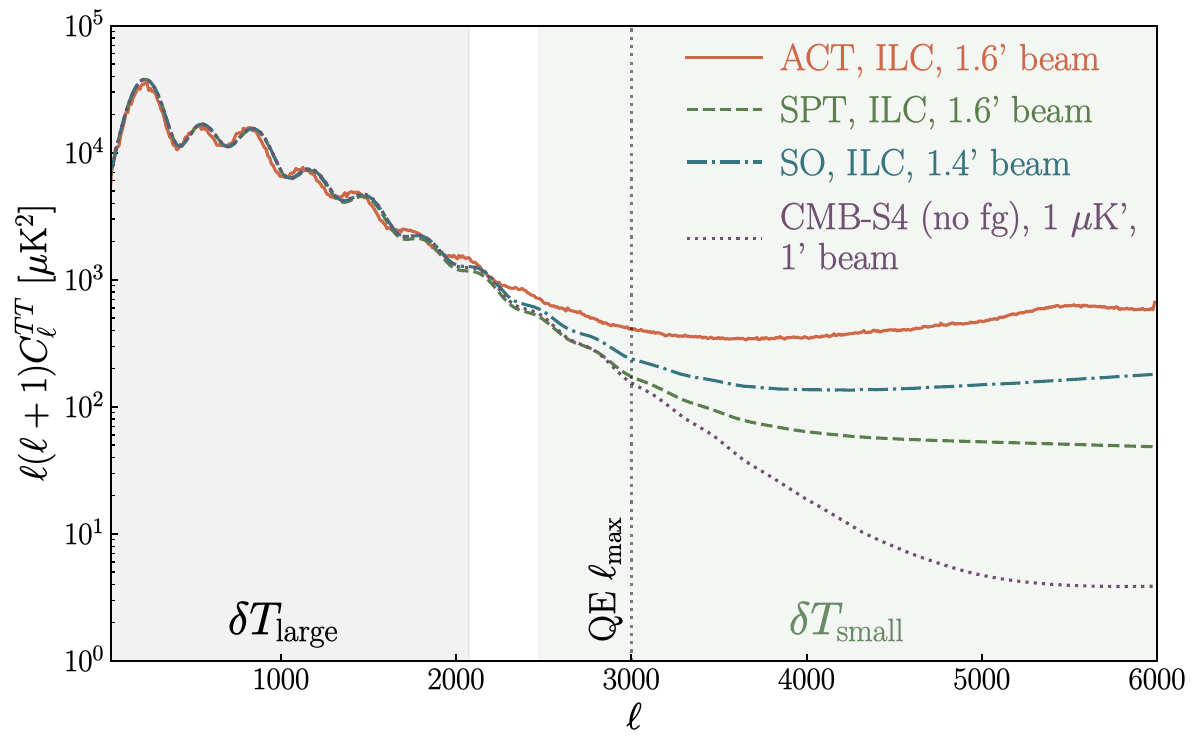}
\caption{Power spectra used in the SNR forecasts for ACT (solid orange), SPT (dashed green), SO (dash-dotted blue), and CMB-S4 (dotted purple).
Details for the spectra are given in App.~\ref{app:tisims}.
The shaded bands show the $\ell$ ranges used by the TI estimator: $30<\ell\lesssim 2075$ (grey) and $2425\lesssim\ell<5950$ (green) for large- and small-scale temperature measurements, respectively.
See Eqs.~\ref{eq:lpf} and \ref{eq:hpf} for the exact $\ell$ filtering used.
The QE uses $30<\ell<3000$ (black dotted line) as it is sensitive to foreground contamination at higher $\ell$.
Since the TI estimator is robust to foregrounds at high $\ell$, it can take advantage of future experiments' lower detector noise and smaller beam widths.
}
\label{fig:tt_power}
\end{figure}

To calculate the covariance matrix, we simulate a set of 128 maps for each CMB experiment, using the power spectra shown in Fig.~\ref{fig:tt_power}.
All of the spectra are beamed and include noise contributions from the CMB, detector noise and--with the exception of CMB-S4--extragalactic foreground contamination (e.g. lensing, CIB, tSZ, kSZ, radio sources).
More details may be found in App. \ref{app:tisims_noise}.

To separate the large- and small-scale temperature fluctuations  $\delta T_L$ and $\delta T_S$, we apply low- and high-pass filters (see Eqs.~\eqref{eq:lpf} and \eqref{eq:hpf}) to generate two separate maps.
We have implemented the stacked TI and signed estimators in the aperture photometry filtering and stacking software, \texttt{ThumbStack} \citep{Schaan2021a}.
We use this code to generate small postage stamps centered on each galaxy position for both maps, apply annular (i.e. ``ring'') aperture photometry filters to measure the $\delta T_S$ radial profile for each stamp, and stack the stamps according to Eq.~\eqref{eq:def_stacked_TI}, using each stamp's average $\delta T_L$ as weights.

We construct the covariance matrix taking into account the clustering of galaxies and scaling by the proper Hartlap correction factor \cite{Hartlap2007}. Details are provided in App. \ref{app:tisims_cov}.

We then measure the SNR for each CMB-galaxy sample combination, scaling to each experiment's full galaxy sample and sky fraction, $f_{\rm sky}$, shown in Tab.~\ref{tab:ngal}.
The resulting SNR values for the surveys are shown in Tab.~\ref{tab:snr}.

For the QE of $\tau$, details of the SNR estimation for the cross-power spectrum of the $\tau$ map reconstructed using QE and galaxy map $g$ are presented in App.~\ref{app:qesnr}. The SNR is calculated analytically as given in Eq.~\eqref{eq:snr} where the noise terms contain the total power spectrum of the reconstructed $\tau$ map along with its reconstruction noise and the total galaxy power spectrum, which includes both the clustering and shot noise contributions. 
The shot noise levels for each galaxy sample are calculated using the number density values provided in Tab.~\ref{tab:ngal}.

\begin{table}[H]
    \centering
    \begin{tabular}{@{}lccc@{}}
        \toprule
        \begin{tabular}{c}Galaxy\\sample\end{tabular} &
        \begin{tabular}{c}$\bar{n}$\\$\rm {arcmin}^{-2}$\end{tabular} &
        \begin{tabular}{c}SPT\\($f_{\rm sky}=0.036$)\end{tabular} &
        \begin{tabular}{c}ACT/SO/CMB-S4\\($f_{\rm sky}=0.315$)\end{tabular} \\
        \midrule
        \begin{tabular}{l}CMASS\end{tabular} & 
        \begin{tabular}{c}0.012\end{tabular} & 
        \begin{tabular}{c}65,600\end{tabular} &
        \begin{tabular}{c}568,776\end{tabular}  \\
        \begin{tabular}{l}unWISE\end{tabular} & 
        \begin{tabular}{c}1.24\end{tabular} & 
        \begin{tabular}{c}6.8M\end{tabular}  & 
        \begin{tabular}{c}59M\end{tabular} \\
        \begin{tabular}{l}LSST\end{tabular} &
        \begin{tabular}{c}34.1\end{tabular} &
        \begin{tabular}{c}187M\end{tabular} & 
        \begin{tabular}{c}1.62B\end{tabular} \\
        \bottomrule
    \end{tabular}
    \caption{Number of galaxies assumed to overlap with each CMB experiment used in the SNR forecasts.
    The shot noise, $\bar{n}$, for each galaxy sample and the sky fraction, $f_{\rm sky}$, for each CMB experiment are also listed. ACT, SO, and CMB-S4 are assumed to have equal $f_{\rm sky}$.
    }
    \label{tab:ngal}
\end{table}

\subsection{Galaxy clustering effects}

As we go towards higher density galaxy samples, the shot noise of the sample decreases by $\sqrt{N_{\rm gal}}$, but the clustering power spectrum still adds noise to the covariance.
Here we assess the impact of clustering effects for the TI and QE forecasts.\\
For the QE, we complete the simulated analysis with and without the galaxy clustering ($gg$) contribution at the power spectrum level, where the  total galaxy power spectrum is calculated with and without the clustering power spectrum. The results of this study are summarized in Tab. \ref{tab:no_gg_snr}.
\begin{table}[H]
    \centering
    \begin{tabular}{@{}llrrr@{}}
    \toprule
    & & \multicolumn{3}{c}{$gg=0$ / $gg\neq0$} \\
    \cmidrule(l{.1em}r{.1em}){3-5}
      \begin{tabular}{l}CMB\\Expt.\rule{1.5em}{0pt}\end{tabular} &
      \begin{tabular}{l}Estimator\rule{0.5em}{0pt}\end{tabular} &
      \begin{tabular}{c}
          CMASS\\
      \end{tabular} &
      \begin{tabular}{c}
          \rule{1em}{0pt}unWISE \\
      \end{tabular} &
      \begin{tabular}{c}
          \rule{1em}{0pt}LSST \\
      \end{tabular} \\
      \toprule
      SPT &
      \begin{tabular}{l}
        TI$_{5950}$ \\
        QE$_{6000}$
      \end{tabular} &
      \begin{tabular}{r}
        0.34 / 0.37 \\
        0.54 / 0.53
      \end{tabular} &
      \begin{tabular}{r}
        3.4 / 3.0 \\
        5.5 / 4.7
      \end{tabular} &
      \begin{tabular}{r}
        18 / 11 \\
        29 / 12
      \end{tabular} \\
      \midrule
      ACT &
      \begin{tabular}{l}
        TI$_{5950}$ \\
        QE$_{6000}$
      \end{tabular} &
      \begin{tabular}{r}
        0.60 / 0.65 \\
        1.1 / 1.1
      \end{tabular} &
      \begin{tabular}{r}
        6.1 / 5.2 \\
        11 / 9.3
      \end{tabular} &
      \begin{tabular}{r}
        31 / 18 \\
        58 / 22
      \end{tabular} \\
      \midrule
      SO &
      \begin{tabular}{l}
        TI$_{5950}$ \\
        QE$_{6000}$
      \end{tabular} &
      \begin{tabular}{r}
        0.84 / 0.92 \\
        1.4 / 1.4
      \end{tabular} &
      \begin{tabular}{r}
        8.5 / 7.5 \\
        14 / 12
      \end{tabular} &
      \begin{tabular}{r}
        45 / 25 \\
        73 / 29
      \end{tabular} \\
      \midrule
      CMB-S4 &
      \begin{tabular}{l}
        TI$_{5950}$ \\
        QE$_{6000}$
      \end{tabular} &
      \begin{tabular}{r}
        1.5 / 1.7 \\
        2.7 / 2.7
      \end{tabular} &
      \begin{tabular}{r}
        16 / 13 \\
        28 / 25
      \end{tabular} &
      \begin{tabular}{r}
        82 / 49 \\
        145 / 65
      \end{tabular} \\
      \bottomrule
    \end{tabular}
    \caption{
    Summary of the detection SNRs for TI and QE varying the noise  contribution due to galaxy clustering, $gg$.
    The first value in each pair corresponds to the case with no clustering ($gg=0$) power spectrum contribution to the total galaxy power spectrum; the second value includes it ($gg\neq0$).
    For QE, we use $60<\ell<6000$.
    For TI, we use $30<\ell\lesssim 2075$ for $\delta T_{L}$ weights and $2425\lesssim\ell<5950$ for $\tau$ reconstruction.
    Ratios of the $gg=0$ and $gg\neq0$ SNR values are visualized in Fig. \ref{fig:snr_ratio}.
    }
    \label{tab:no_gg_snr}
\end{table}

For the TI estimator, we compare the SNR values obtained with a covariance matrix estimated from bootstrapping from a single map realization (thus neglecting clustering effects) versus that estimated from the sample variance over 128 map realizations.

\begin{figure}[H]
\centering
\includegraphics[width= \columnwidth]{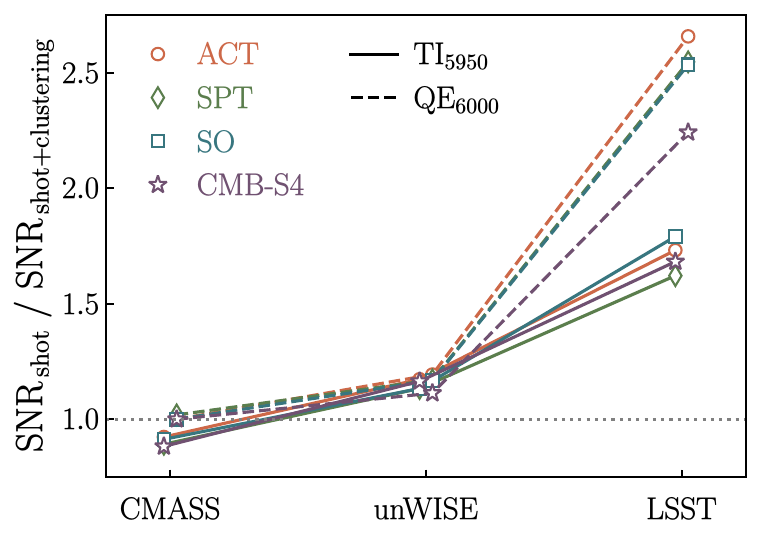}
\caption{Ratio of the SNRs when accounting for galaxy shot noise only versus galaxy shot noise and clustering.
For all three galaxy samples, the difference in SNR ratio between CMB experiments (same colors as Fig. \ref{fig:tt_power}) is relatively small.
However, the impact of clustering on SNR increases significantly for higher density galaxy samples, especially LSST.
When neglecting clustering in the LSST sample, the SNR is overestimated by $\sim60\%$ for TI (solid lines) and $\sim150\%$ for QE (dashed lines).
}
\label{fig:snr_ratio}
\end{figure}

As shown in Fig. \ref{fig:snr_ratio}, neglecting clustering effects results in overestimation of the SNR for both the QE and TI.
While the CMASS sample is sparse enough to not show significant differences, using unWISE results in a moderate $\sim$20\% overestimated SNR.
Ignoring clustering in the LSST sample is a more dramatic oversight: $\sim$60\% for TI and $\sim$150\% for QE.
The marked differences in sensitivity to clustering effects in TI and QE, especially for an LSST-like experiment, is unclear.
We leave an investigation of this difference to future work.
Despite these uncertainties, it is clear clustering has a significant effect on noise estimation.
Thus, we include these effects in the present forecasts and emphasize they must be properly accounted for in future analyses.

We also study the effect of neglecting the cosmic variance of the signal itself, i.e., $C_\ell^{\tau g}$, and find it is negligible for all the experiments considered.

\begin{figure}[h]
\centering
\includegraphics[width= \columnwidth]{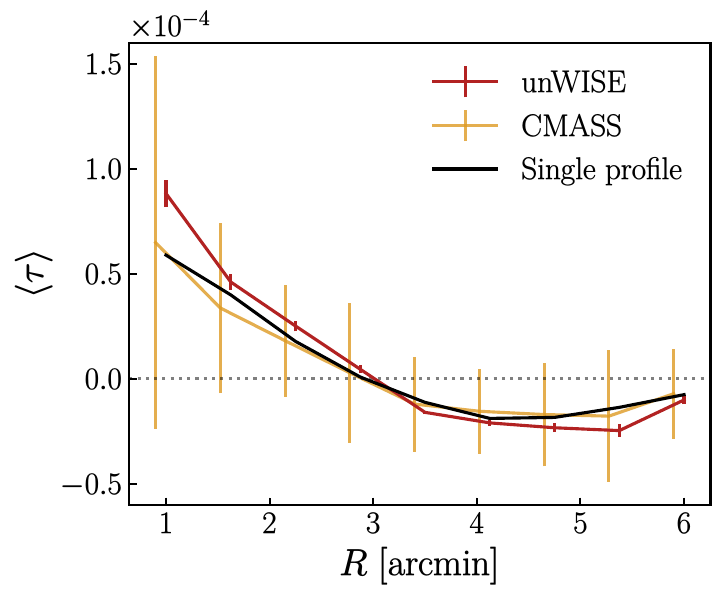}
\caption{Comparison of the high-pass filtered radial $\langle\tau\rangle$ profile when stacking on the unWISE (red) and CMASS (orange) $\tau$ maps with the TI estimator.
The input profile used when painting each galaxy in both maps is shown for comparison (black).
Note the shape of the profiles dips below zero due to the high-pass filtering applied.
The $\tau$ profile stacked on CMASS galaxy locations matches the profile painted on an individual galaxy, i.e., the CMASS sample is sparse enough for the two-halo term to have negligible contribution.
However, after stacking on the unWISE sample, the profile amplitude shows a slight excess, which reflects the two-halo term contribution.
The error bars on the CMASS and unWISE profiles correspond to the expected noise from ACT.
A tentative detection of this two-halo term may be possible with ACT$\times$unWISE.
}
\label{fig:2_halo_profile}
\end{figure}

Figure~\ref{fig:2_halo_profile} shows the comparison of the stacked high-pass filtered radial $\tau$ profile
on the unWISE (orange) and CMASS (red) locations in our simulations where the $\tau$ profile has been reconstructed using the TI esimator with an ACT-like experiment. The input profile used when painting each galaxy in both maps is shown for comparison in black. The average $\tau$ profile stacked on CMASS galaxy locations matches the profile painted on an individual galaxy, i.e., the CMASS sample is sparse enough for the two-halo term to have negligible contribution.
However, after stacking on the unWISE sample, there is a slight excess in the amplitude of the profile, which reflects the two-halo term contribution.
This shows that the clustering term already starts becoming important for the unWISE-like galaxy sample and will be even more important for an LSST-like sample. A tentative detection of this two-halo term might be possible using ACT$\times$unWISE data.

\subsection{Results}

\begin{figure}[H]
\centering
\includegraphics[width= \columnwidth]{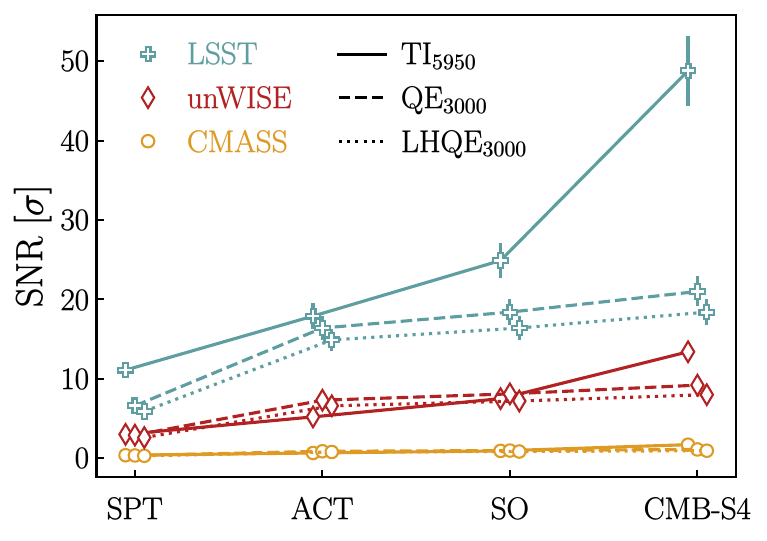}
\caption{Visualization of the forecast SNR values given in Tab. \ref{tab:snr}.
The LHQE (dotted lines) results in a SNR reduction compared to the standard QE (dashed lines).
For current experiments, the QE and TI (solid lines) give similar SNRs.
Cross-correlation with CMASS (orange circles) does not yield a significant measurement for any CMB experiment, whereas the signal is detectable using the unWISE sample (red diamonds).
For future experiments, TI can attain higher SNR than the QE, most dramatically when using the LSST galaxy sample (blue pluses), because it can use smaller CMB scales.}
\label{fig:snr}
\end{figure}

The forecast SNRs for TI, QE, and lens-hardened QE (LHQE), using realistic simulations and accounting for clustering effects as described above, are summarized in Tab. \ref{tab:snr} and visualized in Fig. \ref{fig:snr}.
Unlike for the clustering study, we limit $\lmax$ to 3000 for the QE and LHQE to mitigate foreground contamination.
As TI is robust to foregrounds, we use scales up to $\lmax=5950$.\footnote{This rather specific value corresponds to the most conservative $\lmax$ of the input ILC power spectra used in the simulations.}
As expected, we find that the LHQE degrades the SNR to some extent ($\sim 10-15\%$) when compared to the standard QE as the noise in the reconstruction of $\tau$ for a LHQE goes up. As this is the forecast SNR for cross-correlation of $\tau$ and galaxies $C_\ell^{\tau g}$ rather than auto-power spectrum of $\tau$ $C_\ell^{\tau \tau}$, effects of the noise penalty for hardening against lensing are slightly less dramatic. Factors contributing to the noise (denominator of Eq.~\eqref{eq:snr}) are total galaxy and $\tau$ power spectra. As expected, we find that for a given galaxy sample, as we go towards more sensitive CMB experiments, effects of lens-hardening are felt slightly more as the noise penalty is higher for more sensitive experiments \cite{Roy_2023}. As a result, SNR is affected more by lens-hardening for a CMB-S4-like experiment than an ACT-like experiment. In the following discussion, we limit our comparisons to TI and QE.

For current experiments like ACT and SPT in combination with CMASS and unWISE galaxies, TI and QE give similar results.
Due to fairly high shot noise, the cross-correlation with the CMASS-like sample does not yield a significant SNR for any CMB experiment, whereas the signal is detectable using an unWISE-like sample.
We note this result is consistent with \cite{Coulton_2024}, which uses ACT$\times$unWISE data to make the first detection of this effect.

For future CMB experiments, the TI estimator can attain higher SNR than the QE, most dramatically when using an LSST-like galaxy sample (see Tab.~\ref{tab:snr} and Fig.~\ref{fig:snr}).
This is due to the use of smaller scales of the CMB.
Future CMB experiments, particularly CMB-S4, have exquisite projected sensitivity and beam size.
With lower noise and a smaller beam, the TI estimator can extract information from the small scales that QE cannot reliably use due to foreground biases.
Thus, the SNR values across all cross-correlations with LSST are improved for TI compared to the QE.

\begin{table}[H]
    \centering
    \begin{tabular}{@{}llllll@{}}
    \toprule
      \begin{tabular}{l}Galaxy\\sample\end{tabular} &
      \begin{tabular}{l}Estimator\end{tabular} &
      \begin{tabular}{l}SPT\end{tabular} &
      \begin{tabular}{l}ACT\end{tabular} &
      \begin{tabular}{l}SO\end{tabular} &
      \begin{tabular}{l}CMB-S4\end{tabular}\\
      \toprule
      \begin{tabular}{l}CMASS\end{tabular} &
      \begin{tabular}{l}
        TI$_{5950}$ \\
        QE$_{3000}$\\
        LHQE$_{3000}$ \\
      \end{tabular} &
      \begin{tabular}{l}
        0.37$\sigma$ \\
        0.34$\sigma$ \\
        0.30$\sigma$ \\
      \end{tabular} &
      \begin{tabular}{l}
        0.65$\sigma$ \\
        0.87$\sigma$ \\
        0.78$\sigma$ \\
      \end{tabular} &
      \begin{tabular}{l}
        0.92$\sigma$ \\
        0.96$\sigma$ \\
        0.84$\sigma$ \\
      \end{tabular} &
      \begin{tabular}{l}
        1.7$\sigma$ \\
        1.1$\sigma$ \\
        0.94$\sigma$ \\
      \end{tabular} \\
    \midrule
    \begin{tabular}{l}unWISE\end{tabular} &
      \begin{tabular}{l}
        TI$_{5950}$ \\
        QE$_{3000}$\\
        LHQE$_{3000}$ \\
      \end{tabular} &
      \begin{tabular}{l}
        3.0$\sigma$ \\
        2.9$\sigma$ \\
        2.6$\sigma$ \\
      \end{tabular} &
      \begin{tabular}{l}
        5.2$\sigma$ \\
        7.3$\sigma$ \\
        6.6$\sigma$ \\
      \end{tabular} &
      \begin{tabular}{l}
        7.5$\sigma$ \\
        8.1$\sigma$ \\
        7.2$\sigma$ \\
      \end{tabular} &
      \begin{tabular}{l}
        13$\sigma$ \\
        9.2$\sigma$ \\
        8.0$\sigma$ \\
      \end{tabular} \\
    \midrule
      \begin{tabular}{l}LSST\end{tabular} &
      \begin{tabular}{l}
        TI$_{5950}$ \\
        QE$_{3000}$\\
        LHQE$_{3000}$ \\
      \end{tabular} &
      \begin{tabular}{l}
        11$\sigma$ \\
        6.6$\sigma$ \\
        5.9$\sigma$ \\
      \end{tabular} &
      \begin{tabular}{l}
        18$\sigma$ \\
        16$\sigma$ \\
        15$\sigma$ \\
      \end{tabular} &
      \begin{tabular}{l}
        25$\sigma$ \\
        18$\sigma$ \\
        16$\sigma$ \\
      \end{tabular} &
      \begin{tabular}{l}
        49$\sigma$ \\
        21$\sigma$ \\
        18$\sigma$ \\
      \end{tabular} \\
    \bottomrule
    \end{tabular}
    \caption{Summary of the forecast detection SNRs for TI and QE accounting for galaxy clustering and using realistic $\ell$ ranges for each estimator. We do not include cosmic variance on the $\tau g$ signal, which is negligible for all experiments considered. For the standard QE and lens-hardened QE (LHQE), we use $60<\ell<3000$ where these estimators have limited response to foreground contamination. For TI, we use the same $\ell$ range as in Tab. \ref{tab:no_gg_snr}. TI can use higher $\ell$ modes as it is robust to foregrounds; the SNR for TI can thus exceed QE because of this added information. All SNR values are visualized in Fig. \ref{fig:snr}.
    }
    \label{tab:snr}
\end{table}

\section{Conclusions}
\label{sec:conclusions}

In this paper, we have derived and studied new estimators for the patchy screening of the CMB, called TI and signed estimators.
These estimators are modifications of the standard screening QE, analogous to the gradient inversion estimators for CMB lensing.

The TI and signed estimators make use of a unique feature of patchy screening: it is the only effect which changes sign based on the local large-scale temperature.
This allows TI and signed estimators to be automatically unbiased to extragalactic foregrounds like the CIB, tSZ and kSZ, unlike the standard QE.
This foreground robustness can be further enhanced for the signed estimator via ``sign balancing'' and ``sign thresholding,'' which further immunise it against foregrounds at a negligible noise cost.
For the QE, the corresponding biases can be large and likely require some form of hardening at some cost in SNR.

The TI and signed estimators are also robust to CMB lensing bias, unlike the standard QE.
Lens hardening remedies this issue for QE, again, with higher noise penalty.

We also clarify the relationship between the TI estimator, signed estimator and QE.
We show that the QE reconstructs both small- and large-scale screening information and present simple intuition about these two regimes.
We show how TI is a subtle modification to the small-scale-only QE, which in principle avoids the noise floor above.
Finally, the signed estimator is a modification to TI which allows even more robustness to foregrounds (via sign balancing and sign thresholding), at the cost of a factor of $\sqrt{\pi/2} \approx 1.25$ reduction in SNR. Even in the absence of detector noise, foregrounds, and lensing, the $\tau$ QE has an irreducible source of variance and hits a noise floor, whereas the TI and signed estimators could perform arbitrarily better under such noiseless conditions. However, this noise floor is not reached in practice.
For a fixed set of Fourier scales included in the analysis, the statistical SNR in the TI or signed estimators are similar to the QE but do not outperform it.
However, by being robust to lensing and foregrounds, TI and signed estimators can be applied to more Fourier modes (pushing to smaller scales), thus potentially increasing the overall SNR.

We perform realistic forecasts for current and upcoming CMB and LSS experiments (see Tab.~\ref{tab:snr} and Fig.~\ref{fig:snr}).
Our results highlight the large noise enhancement due to the clustering of galaxies once the galaxy number density reaches LSST-like values.
Further, our SNR forecast for ACT$\times$unWISE is consistent with the measurement found in the companion paper \cite{Coulton_2024}, which is the first detection of this signal.

We have focused on measuring the patchy screening from the CMB temperature only.
The TI and signed estimators should work identically when substituting temperature T with Stokes Q or U polarization maps.
We leave this exploration to future work.

In the future, patchy screening measurements around galaxies will be a powerful probe of the gas density profile around them, highly complementary with kSZ and tSZ effects.
Joint analyses of these signals around the same galaxies will yield valuable insights into the thermodynamical properties of baryons within these halos and shed more light on the lesser understood properties of feedback processes. This will have a direct impact on weak galaxy lensing studies where baryonic effects are a major systematic.
They will also help in breaking the velocity-optical depth degeneracy in kSZ, thus expanding its power as a probe of the large-scale velocities to constrain the growth rate of structure, modifications to general relativity and local primordial non-Gaussianity.
In particular, as the uncertainty on the kSZ ($\propto \tau v_\text{LOS}$) measurements will be negligible compared to that of patchy screening ($\propto \tau$), the SNR values presented here translate as the expected measurement SNR on the growth rate of structure ($\propto v_\text{LOS}$).

\section*{Data Availability}
The \texttt{LensQuEst} and \texttt{ThumbStack} software packages are publicly available under an open source license.

\section*{Acknowledgments}

We thank Nicholas Battaglia, Simone Ferraro, Darby Kramer, Srinivasan Raghunathan, Rashid Sunyaev, Alexander van Engelen and Ziang Yan for useful discussions.
We thank Joanna Dunkley, Lyman Page and Edward Wollack for helpful comments on the manuscript.

This material is based upon work supported by the National Science Foundation Graduate Research Fellowship under Grant No. DGE-2146755.
TS thanks the LSSTC Data Science Fellowship Program, which is funded by LSSTC, NSF Cybertraining Grant \#1829740, the Brinson Foundation, and the Moore Foundation; their participation in the program has benefited this work.
This work received support from the U.S. Department of Energy under contract number DE-AC02-76SF00515 to SLAC National Accelerator Laboratory.
Some of the computing for this project was performed on the Sherlock cluster. We thank Stanford University and the Stanford Research Computing Center for providing computational resources and support.

\bibliographystyle{prsty.bst}
\bibliography{refs, erratum_refs}

\appendix

\section{Optimal $\delta T_L$ estimator in the presence of foreground bias}
\label{app:optimal_tlarge_estimator}

In this section, we derive estimators of the large-scale CMB temperature around a given galaxy.
To declutter the derivation, we simplify our notation as:
\beq
d_\vl = T_\vl + n_\vl + A u_\vl,
\eeq
where $d_\vl$ is the temperature data as a function of Fourier multipole $\vl$, $T_\vl$ is the true CMB (with known power spectrum $C_\ell$) to be estimated, $n_\vl$ is the noise (with known power spectrum $N_\ell$), and $A u_\vl$ is a foreground emission, with a known profile $u_\vl$ (e.g., the beam for a point source) and unknown amplitude $A$.
We would like to build an estimator $\hat{T}_\vl$ of $T_\vl$, which minimizes the mean squared error 
$\langle \left( \hat{T}_\vl - T_\vl \right)^2 \rangle$.
We restrict ourselves to estimators linear in the data, i.e. we wish to determine the weights $W_\vL$ for each $\vl$ such that:
\beq
\hat{T}_\vl
=
\sum_\vL W_\vL d_{\vl-\vL}.
\eeq

In the absence of foregrounds, i.e., if $A$ is known to be zero, minimizing the mean-squared error leads to the well-known Wiener filter:
\beq
\hat{T}_\vl
=
\frac{C_\ell}{C_\ell + N_\ell} d_\vl.
\eeq
The Wiener filter can also be derived as the maximum \textit{a posteriori} estimator for $T_\vl$, in the absence of foregrounds ($A=0$), as well as in the presence of foregrounds for which the prior on the amplitude $A$ is flat.
Such foregrounds would have a random sign, which is typically not the case in our problem.

We now consider the case of non-zero foreground and impose the additional constraint that the estimator $\hat{T}_\vl$ should have zero response to the foreground amplitude:
\beq
\frac{d\hat{T}_\vl}{dA} = 0.
\eeq
In what follows, we show that the linear estimator which minimizes the mean squared error, while having zero response to the foreground emission, is simply obtained by Wiener-filtering the data, after having subtracted the matched-filtered foreground template.

We introduce a Lagrange multiplier $\lambda$, to minimize the mean-squared error under the constraint
\beq
\mathcal{L}
=
\langle \left( \hat{T}_\vl - T_\vl \right)^2 \rangle
-
\lambda \frac{d\hat{T}_\vl}{dA}.
\eeq
In terms of the weights $W_\vL$, this loss function becomes
\beq
\bal
\mathcal{L}
&=
\left( W_0-1 \right)^2 C_\ell
+
W_0^2 N_\ell \\
&+ \sum_{\vL\neq 0} W_\vL^2 \left( C_{\vl-\vL} + N_{\vl-\vL} \right)\\
&+ A^2 \sum_{\vL, \vL^\prime} W_\vL W_{\vL^\prime} u_{\vl-\vL} u_{\vl-\vL^\prime}\\
&- \lambda \sum_{\vL} W_\vL u_{\vl-\vL}.
\eal
\label{eq:WL_lambda}
\eeq
Setting $\partial \mathcal{L}/\partial W_\vL$ to zero gives
\beq
W_\vL =
\frac{C_\ell}{C_\ell + N_\ell} \delta^K_{\vL, 0}
-
\left( \alpha + \lambda \right)
\frac{u_{\vl-\vL}}{C_{\vl-\vL} + N_{\vl-\vL}},
\eeq
where we have defined
\beq
\alpha 
\equiv
A \sum_{\vL^\prime}
W_{\vL^\prime} u_{\vl - \vL^\prime}.
\eeq
Setting $\partial \mathcal{L}/\partial \lambda$ to zero gives $\sum_{\vL}
W_{\vL} u_{\vl - \vL} = 0$,
which means that $\alpha=0$ and lets us solve for $\lambda$ in Eq.~\eqref{eq:WL_lambda}:
\beq
\lambda
=
\frac{C_\ell}{C_\ell+N_\ell} \
\frac{u_\vl}
{\left(\sum_\vL \frac{u_\vL^2}{C_L + N_L}\right)}.
\eeq
Plugging this into Eq.~\eqref{eq:WL_lambda} yields the expressions for the weights:
\beq
W_\vL
=
\frac{C_\ell}{C_\ell + N_\ell}
\left[ 
\delta^K_{\vL, 0}
-
\frac{\frac{u_{\vl-\vL}}{C_{\vl-\vL} + N_{\vl-\vL}}}
{\left( \sum_{\vL^\prime} \frac{u_{\vL^{\prime 2}}}{C_{L^\prime} + N_{L^\prime}} \right)}
u_\vl
\right].
\eeq
In conclusion, the estimator for $T_\vl$, linear in the 
data $d_\vl$, which minimizes the mean squared error while having zero response to the foreground emission is 
\beq
\hat{T}_\vl
=
\frac{C_\ell}{C_\ell + N_\ell}
\left[
d_\vl
-
\hat{A} u_\vl
\right],
\eeq
where $\hat{A}$ is the matched filter estimator for the foreground amplitude $A$:
\beq
\hat{A}
\equiv
\frac{\left(\sum_{\vL}\frac{d_{\vL}u_{\vL}}{C_{L} + N_{L}}\right)}
{\left( \sum_{\vL^\prime} \frac{u_{\vL^{\prime 2}}}{C_{L^\prime} + N_{L^\prime}} \right)}.
\eeq
As stated, this reduces to subtracting the foreground template $\hat{A}u_\vl$ from the data, where $\hat{A}$ is the matched filter for the profile $u_\vl$, and then Wiener-filtering the result.

\section{Variance comparison: small-scale QE vs. TI}
\label{app:variance_comparison_ti_qe}

We wish to get intuition about the noise properties of the QE and TI,
in order to understand in which case one may be a better estimator than the other.

We decompose the observed small and large-scale temperatures around a given galaxy into signal and noise:
\beq
\left\{
\bal
&\delta \hat{T}_S = - \delta T_L \tau + n_S\\
&\delta \hat{T}_L = \delta T_L + n_L\\
\eal
\right.
.
\label{eq:observed_vs_true_temperatures}
\eeq
Here, the noise $n_S$ includes everything in the small-scale map other than our screening effect, i.e. detector noise, foregrounds, lensing and the small, Silk-damped primary CMB.
The large-scale noise $n_L$ should be thought of as very small, since the primary CMB is very well-measured on degree scales.

To get a sense of the relevant noise regimes, we introduce the variances
$\sigma^2_{T_L}, \sigma^2_\tau, \sigma^2_{n_L}$ and $\sigma^2_{n_S}$
of $\delta \hat{T}_L$, the true $\tau$, $n_L$ and $n_S$.
In what follows, we assume $\sigma^2_{n_L} \ll \sigma^2_{T_L}$, realistic for current and upcoming experiments,
and that $\delta T_L, \tau, n_L$ and $n_S$ are statistically independent.

We can now express the TI and QE estimators in real space as:
\beq
\left\{
\bal
&\hat{\tau}^\text{TI} = - \delta \hat{T}_S / \delta \hat{T}_L\\
&\hat{\tau}^\text{QE} = - \delta \hat{T}_S \delta \hat{T}_L / \sigma^2_{T_L}\\
\eal
\right.
\label{eq:response_noise_TI_QE}
\eeq

We then plug in the expressions for the observed quantities Eq.~\eqref{eq:observed_vs_true_temperatures}, in terms of the true ones, into the QE and TI estimators.
As expected, we obtain a linear response in the true $\tau$ and a noise term:
\beq
\left\{
\bal
&\hat{\tau}^\text{TI} = \alpha^\text{TI} \tau +n^\text{TI}\\
&\hat{\tau}^\text{QE} = \alpha^\text{QE} \tau +n^\text{QE}\\
\eal
\right.
.
\eeq

\subsection{Irreducible noise in the QE response}

For the TI estimator, 
\beq
\alpha^\text{TI}
=
\frac{\delta T_L}{ \delta T_L + n_L}
=
1 - \frac{n_L}{\delta T_L} + \left(\frac{n_L}{\delta T_L}\right)^2 +...
\eeq
such that the mean response $\langle \alpha^\text{TI} \rangle$ differs from unity by a bias $\mathcal{O}\left( \sigma^2_{n_L} / \sigma^2_{T_L} \right)$.
The variance of the response $\alpha^\text{TI}$ is similarly small, with
\beq
\text{var}\left( \alpha^\text{TI} \right) 
= \mathcal{O}\left( \sigma^2_{n_L} / \sigma^2_{T_L} \right)
.
\eeq
Thus, in practice, the bias and the variance of the TI response are negligibly small, i.e. $\alpha^\text{TI} \simeq 1$.

In contrast, the linear response of the QE estimator is
\beq
\alpha^\text{QE}
=
\frac{\delta T_L \left(\delta T_L + n_L \right)}{\sigma^2_{T_L}}
.
\eeq
This response is exactly unbiased, i.e. $\langle \alpha^\text{QE} \rangle = 1$.
However, its variance does not go to zero in the limit of low noise:
\beq
\text{var}\left( \alpha^\text{TI} \right) 
= 
2 + \sigma^2_{n_L} / \sigma^2_{T_L}.
\eeq
Thus, if the noises $n^\text{TI}$ and $n^\text{QE}$ are negligible, 
the TI estimator converges to a perfect reconstruction of the true $\tau$,
whereas the QE estimator converges to $\alpha^\text{QE} \tau$, where $\alpha^\text{QE}$ is a random number with variance 2.

\subsection{Comparable estimator noises}

The noise $n^\text{TI}$ of the TI and QE estimators are
\beq
\left\{
\bal
&n^\text{TI}
=
\frac{n_S}{\delta T_L + n_L} \\
&n^\text{QE}
=
\frac{n_S \left( \delta T_L + n+L \right)}{\sigma^2_{T_L}} \\
\eal
\right.
.
\eeq
Since
$\langle 1/\delta T_L^2 \rangle \gtrsim 1/\sigma^2_{T_L}$ by convexity of $x \rightarrow 1/x$,
we obtain
\beq
\text{var}\left( n^\text{TI} \right)
\gtrsim 
\frac{\sigma^2_{n_S}}{\sigma^2_{T_L}}
\left( 1 + \frac{\sigma^2_{n_L}}{\sigma^2_{T_L}} \right)
=
\text{var}\left( n^\text{QE} \right).
\eeq
Thus, while the TI noise is larger than the QE noise, they are comparable.

\subsection{Total variance: TI outperforms QE in the low-noise regime}

To decide in which regime one estimator might be preferred over the other, we take into account all the contributions to the estimator variance, including the true signal $\tau$, the noise $n^\text{TI/QE}$ and the response $\alpha^\text{TI/QE}$.
This leads to 
\beq
\text{var}\left( \hat{\tau}^\text{TI/QE} \right)
=
\text{var}\left( \alpha^\text{TI/QE} \right)
\sigma^2_\tau
+
\text{var}\left( n^\text{TI/QE} \right),
\eeq
such that we find:
\beq
\left\{
\bal
&\text{var}\left( \hat{\tau}^\text{TI} \right)
=
\frac{\sigma^2_{n_L}}{\sigma^2_{T_L}}\ \sigma^2_\tau
+
\mathcal{O}\left( \frac{\sigma^2_{n_S}}{\sigma^2_{T_L}} \right) \\
&\text{var}\left( \hat{\tau}^\text{QE} \right)
=
\left( 2 + \frac{\sigma^2_{n_L}}{\sigma^2_{T_L}}\right)\ \sigma^2_\tau
+
\frac{\sigma^2_{n_S}}{\sigma^2_{T_L}} \\
\eal
\right.
\eeq
It thus appears that the TI is preferred over the QE when the ``2'' term is relevant, i.e. when:
\beq
\sigma^2_{n_L} \lesssim \sigma^2_{T_L}
\text{  and  }
\sigma^2_{n_S} \lesssim \sigma^2_{T_L} \sigma^2_\tau.
\eeq
The first condition is easily satisfied, since the large scale temperature is measured at high SNR in current experiments.
The second condition effectively requires that the $n_S$ be smaller than the screening effect $\delta T_L \tau$ in the measured small-scale temperature.
However, the noise $n_S$ includes not only detector noise, which could in principle be made arbitrarily small, but also foregrounds and lensing.
Thus, going beyond QE with TI requires significant foreground cleaning.
Lensing is much larger than screening in the small-scale temperature power spectrum, which violates
$\sigma^2_{n_S} \lesssim \sigma^2_{T_L} \sigma^2_\tau$.
Thus, the TI SNR should only outperform the QE's if a large amount of delensing can be achieved.
We leave quantifying this required amount of delensing to future work.

\section{Two versions of the approximate small-scale QE; same comparison with TI}
\label{app:stacked_TI_vs_QE}

Here we give the expressions for the stacked TI and QE estimators.
These show that the key difference between TI and QE remains true for the stacked estimators: the QE's response to the true $\tau$ has an irreducible variance, even in the absence of any noise.

As we show in the main text, the individual TI estimator for galaxy $i$ is
$\tau^\text{TI}_i \equiv \delta\hat{T}_{Si} / \delta\hat{T}_{Li}$,
with variance $\sigma^2_{n_s} / \delta\hat{T}_{Li}^2$,
such that the inverse-variance weighted average over all galaxies is given by Eq.~\eqref{eq:def_stacked_TI}:
\beq
\hat{\tau}^\text{TI stack}
=
\frac{
\sum_i
\hat{\delta T}_{S i}\hat{\delta T_{Li}} / \sigma_{n_S}^2
}
{
\sum_j \hat{\delta T_L}_j^2 / \sigma_{n_S}^2
}
.
\eeq
Ignoring the small large-scale noise $n_L$, but taking into account the small-scale noise $n_S$, this can be rewritten as
\beq
\hat{\tau}^\text{TI stack}
=
\underbrace{1}_\text{TI response} 
\times \tau
+
\underbrace{
\frac{
\sum_i
n_{S i}\hat{\delta T_{Li}} / \sigma_{n_S}^2
}
{
\sum_j \hat{\delta T_L}_j^2 / \sigma_{n_S}^2
}
}_{\text{TI noise}}
,
\eeq
i.e. the response to the true $\tau$ is unity (again, for small $n_L$).

On the other hand, the small-scale QE estimator for a given galaxy $i$ takes the form
$\tau^\text{QE}_i \equiv \delta\hat{T}_{Si} \delta\hat{T}_{Li} / \sigma^2_{T_L}$.
When quantifying the variance of $\tau^\text{QE}_i$, two options are possible.
A ``local'' variance estimate takes into account the local value of $\delta\hat{T}_{Li}$, whereas a ``global'' variance estimate marginalizes over $\delta\hat{T}_{Li}$, given its variance $\sigma^2_{T_L}$.

The global variance estimate $\sigma^2_{n_S} / \sigma^2_{T_L}$ is in the spirit of the standard QE, as it preserves the quadratic nature of the estimator, by not including the data itself in the weighting.
In this case, the stacked estimator is then:
\beq
\bal
\hat{\tau}^\text{QE ``global'' stack}
&=
\frac{
\sum_i
\delta\hat{T}_{Si} \delta\hat{T}_{Li} / \sigma^2_{n_S}
}
{
\sum_j \sigma^2_{T_L} / \sigma_{n_S}^2
}\\
&=
\underbrace{
\frac{
\sum_i
\delta\hat{T}_{Li}^2 / \sigma^2_{n_S}
}
{
\sum_j \sigma^2_{T_L} / \sigma_{n_S}^2
}
}_{\text{response, var}\neq 0}
\times \tau
+
\underbrace{
\frac{
\sum_i
n_{S i}
\delta\hat{T}_{Li} / \sigma^2_{n_S}
}
{
\sum_j \sigma^2_{T_L} / \sigma_{n_S}^2
}
}_\text{noise}
.
\eal
\eeq

If one were to instead weigh by the local variance $\sigma^2_{n_S} / \sigma^4_{T_L} \delta \hat{T}_{Li}$, one obtains the ``local'' stacked QE, no longer quadratic in the data:
\beq
\bal
\hat{\tau}^\text{QE ``local'' stack}
&=
\frac{
\sum_i
\frac{\delta\hat{T}_{Si}}{\delta\hat{T}_{Li}}
\frac{\sigma^2_{T_L}}{\sigma^2_{n_S}}
}
{
\sum_j
\frac{1}{\delta\hat{T}^2_{Lj}}
\frac{\sigma^4_{T_L}}{\sigma^2_{n_S}}
}\\
&=
\underbrace{
\frac{
\sum_i
\sigma^2_{T_L} / \sigma^2_{n_S}
}
{
\sum_j 
\frac{\sigma^2_{T_L}}{\delta\hat{T}_{Lj}^2}
\frac{\sigma^2_{T_L}}{\sigma^2_{n_S}}
}
}_{\text{response, var}\neq 0}
\times \tau
+
\underbrace{
\frac{
\sum_i
\frac{n_{Si}}{\delta\hat{T}_{Li}}
\frac{\sigma^2_{T_L}}{\sigma^2_{n_S}}
}
{
\frac{\sigma^2_{T_L}}{\delta\hat{T}_{Lj}^2}
\frac{\sigma^2_{T_L}}{\sigma^2_{n_S}}
}
}_\text{noise}
.
\eal
\eeq

Thus, both for the ``local'' and ``global'' weightings, the stacked QE estimator differs from the stacked TI estimator.
Furthermore, the response to the true $\tau$ of the stacked QE estimators still has an irreducible variance, even as the small-scale noise $\sigma^2_{n_S}$ goes to zero.

As discussed in the main text, the stacked global QE and stacked TI appear very similar, differing only in their denominator.
The fractional difference between the denominators scales as $1/\sqrt{N_\text{galaxies}}$, such that it becomes smaller and smaller with larger galaxy samples.
However, this difference is not irrelevant, because the variance of the estimators (stacked QE and TI) also decreases as $1/\sqrt{N_\text{galaxies}}$.
Thus, this difference is actually significant.

\section{$\tau^{\rm sgn}$ estimator: Sign balancing helps with imperfect foreground cancellation}
\label{app:sign_balancing_foreground_cancellation}

``Sign balancing'' refers to requiring that the number of positively and negatively weighted galaxies in the stack is the same. Sign balancing is effective as long as the contribution to the foreground emission comes from $>0.1\%$ of the total number of galaxies. This appendix presents the evidence for this statement.\\
The signed estimator $\hat{\tau}^{\rm sgn}$ for the foreground emission becomes
\beq
\hat{\tau}^{\rm sgn} \equiv \frac{\sum_i\sgn(T_{L,i})f_i}{\sum_i|T_{L,i}|}
\eeq
where $T_{L,i}$ is the large-scale temperature at the position of galaxy $i$, and $f_i$ is the small-scale temperature due to foreground emission from galaxy $i$.
Simplifying notation, we write
\beq
\bal
\hat{\tau}^{\rm sgn}
&= \frac{\frac{1}{n_{\rm gal}}\sum_i\sgn(T_{L,i})f_i}{\frac{1}{n_{\rm gal}}\sum_i|T_{L,i}|}\\
&\equiv \frac{T^{\rm N}}{T^{\rm D}},
\eal
\eeq
where $n_{\rm gal}$ is the number of galaxies in the stack, and $T^{\rm N}$ and $T^{\rm D}$ are defined as the numerator and denominator of the r.h.s., respectively.
We have the following relations:
\beq
\left\{
\bal
\sgn(T_{L,i}) &\ind |T_{L,j}|\\
\sgn(T_{L,i}) &\ind f_j\\
|T_{L,i}| &\ind f_j\\
|T_{L,i}| &\nind |T_{L,j}|
\eal
\right.\forall\,i,j,
\eeq
where $\ind$ represents statistical independence of the l.h.s. and r.h.s.. The last relation comes from the fact that galaxies may be located in a region where the CMB temperature is highly correlated (i.e. in $\sim$1 deg$^2$ patches).
Since we assume $\sgn(T_{L,i}) \ind f_j$, there is no mean bias from foregrounds, i.e. $\left<\hat{\tau}\right>=0$.
However, for a given realization we may find $\left<\hat\tau\right>\neq 0$ due to imperfect cancellation among the $n_{\rm gal}$ sources.
Here we derive the typical value of $\taubias$, i.e. the RMS error of $\taubias$, to estimate the bias due to imperfect cancellation.
First, we simplify our expression for $\var(\taubias)$:
\beq
\bal
\var(\taubias)
&= \var\left(\frac{1}{T^{\rm D}}T^{\rm N}\right)\\
&= \var\left(\frac{1}{T^{\rm D}}\right)\var(T^{\rm N}) + \left<\frac{1}{T^{\rm D}}\right>^2\var(T^{\rm N})\\
&\qquad + \underbrace{\left<T^{\rm N}\right>^2}_{=0} \var\left(\frac{1}{T^{\rm D}}\right)\\
&= \left(\var\left(\frac{1}{T^{\rm D}}\right) + \left<\frac{1}{T^{\rm D}}\right>^2\right)\var(T^{\rm N})\\
&= \left<\frac{1}{(T^{\rm D})^2}\right>\var(T^{\rm N}).
\eal
\eeq

Since $\var(\taubias) \propto \var(T^{\rm N})$, we can simply use the variance of the numerator $T^{\rm N}\equiv\frac{1}{n_{\rm gal}}\sum\sgn(T_{L,i})f_i$ for the remainder of the analysis.
We first determine the typical value of $T^{\rm N}$ in the absence of fluctuations in foreground contamination, i.e., $f_i=\left<f_i\right> = f~\forall\,i$.
Within a correlated $\sim1$\,deg$^2$ patch of the CMB, sgn($T_{L,i}$) are identical.
For $n_{\rm patch}$ patches, we expect $\sqrt{n_{\rm patch}}$ excess patches of one sign (positive or negative).
Hence, the total excess number of galaxies with positive or negative weight is
\beq
n_{\rm gal,excess} = \sqrt{n_{\rm patch}}\left(\frac{n_{\rm gal}}{n_{\rm patch}}\right)=\frac{n_{\rm gal}}{\sqrt{n_{\rm patch}}}.
\eeq
Therefore, the typical value of $T^{\rm N}$ is
\beq
\bal
T^{\rm N}_{\rm typical}
&= \frac{1}{n_{\rm gal}}\left(n_{\rm gal,excess}f\right)\\
&= \frac{1}{n_{\rm gal}}\left(\frac{n_{\rm gal}f}{\sqrt{n_{\rm patch}}}\right)\\
&= \frac{f}{\sqrt{n_{\rm patch}}}.
\eal
\eeq
By ``sign balancing,'' i.e., ensuring $n_{\rm gal,excess}=0$ before stacking the sources, the bias due to imperfect cancellation of sources is completely negated in the regime where all sources have equal foreground emission.\\
We can relax this assumption and allow $f_i$ to be distinct for all $i$.
We define $\sigma_f^2 \equiv \var(f_i)$ and assume the source weights are already sign balanced such that $\sum_i\sgn(T_{L,i})=0$.
In this case,
\beq
\bal
{\rm var}(T^{\rm N})
&= \frac{1}{n_{\rm gal}^2} \sum_i\sgn(T_{L,i})^2\sigma_f^2\\
&= \frac{\sigma_f^2}{n_{\rm gal}}.
\eal
\eeq
Thus,
\beq
T^{\rm N}_{\rm typical} = \frac{\sigma_f}{\sqrt{n_{\rm gal}}}.
\eeq
We see sign balancing does not fully negate the imperfect cancellation caused by foreground fluctuations.
It is therefore only effective if the bias due to mismatched positive/negative weights is much greater than the bias due to scatter in foreground levels, i.e.,
\beq
\bal
\label{eq:sgn_ineq}
\frac{f}{\sqrt{n_{\rm patch}}}
&\gg \frac{\sigma_f}{\sqrt{n_{\rm gal}}}\\
\Rightarrow \frac{\sigma_f}{f}
&\ll \sqrt{\frac{n_{\rm gal}}{n_{\rm patch}}}\\
&\ll \sqrt{\rho_{\rm gal}A_{\rm patch}},
\eal
\eeq
where $\rho_{\rm gal}$ is the galaxy number density and $A_{\rm patch}$ is the area of the correlated temperature patch.
$A_{\rm patch}\sim1\,$deg$^2$, and $\rho_{\rm gal}\sim1000\,$deg$^{-2}$ for current galaxy surveys.
Eq. \ref{eq:sgn_ineq} is not satisfied if a small fraction of the galaxy sample dominates the foreground signal.
We use the following toy model to determine the minimum fraction of foreground emitting galaxies required for sign balancing to be effective.
Consider the case where a fraction $x$ of the galaxy sample emits foregrounds, and $1-x$ emits no foregrounds.
The population of emitting galaxies must contribute a total of $f/x$ foreground emission such that $\left<f_i\right> = (f/x)*x + (0)*(1-x) = f$.
The variance of the foreground emission is
\beq
\bal
{\rm var}(f_i)
&= \left<f_i^2\right> - \left<f_i\right>^2\\
&= \left[\left(\frac{f}{x}\right)^2x + 0^2(1-x)\right] - f^2\\
&= f^2\left(\frac{1}{x}-1\right).
\eal
\eeq
Thus,
\beq
\frac{\sigma_f}{f} = \sqrt{\frac{1}{x} - 1}.
\eeq
Substituting this result into Eq. \ref{eq:sgn_ineq} and squaring both sides, we determine
\beq
\bal
\frac{1}{x} - 1
&\ll \rho_{\rm gal}A_{\rm patch}\\
&\ll 1000\\
\Rightarrow x &\gg 10^{-3}.
\eal
\eeq
Thus, sign balancing is effective as long as the foreground emission comes from $>0.1\%$ of the total galaxies.
To ensure this condition is satisfied, one can exclude clusters or high-mass groups from the galaxy sample, which would dominate the foreground contamination.

\section{Mean foreground contamination comparison for stacked TI and signed estimators}
\label{app:foreground_bias_ti_sign}

There are three regimes of foreground contamination for the TI and signed estimators: (1) present small-scale foreground $f_S$ but negligible large-scale foreground $f_L$, (2) present large-scale foreground $f_L$ but negligible small-scale foreground $f_S$, and (3) non-negligible correlated foregrounds $f_S$ and $f_L$.
In the first two cases, $f_S$ (or $f_L$) is uncorrelated with the temperature fluctuation $\delta T_L$ (or $\delta T_S$) and thus introduces no bias to either the TI or signed estimators.
We study the third regime for both estimators below.\\
The stacked TI estimator including both large- and small-scale foreground contamination is
\beq
\hat{\tau}^{\rm TI}
=
-\frac{\sum (\delta T_L + f_L)(\delta T_S + f_S)}{\sum (\delta T_L + f_L)^2}.
\eeq
We assume $f_L \ll \sigma_{T_L}$ (for the typical values of $\sigma_{T_L}\sim100\,\mu$K and $f_L\sim1\,\mu$K this is quite accurate).
For simplicity, we also assume all objects have equal $f_L$ and $f_S$.
The bias to the stacked TI estimator as a function of radial distance $r$ in the 1D stacked photometry profile to first order in $f_L/\sigma_{T_L}$ is
\beq
\bal
\hat{\tau}_{\rm bias}^{\rm TI}(r)
&=
-\left<
\frac{\sum f_S(r) f_L}
{\sum \delta T_L^2}  
\right>
+ \mathcal{O}\left( \left(\frac{f_L}{\sigma_{T_L}}\right)^2 \right)\\
&=
- f_S(r) f_L
\left<
\frac{N_\text{object}}
{\sum \delta T_L^2}  
\right>
+ \mathcal{O}\left( \left(\frac{f_L}{\sigma_{T_L}}\right)^2 \right)\\
&\simeq
- f_S(r) f_L
\frac{1}
{\left<\delta T_L^2\right> }   
+ \mathcal{O}\left( \left(\frac{f_L}{\sigma_{T_L}}\right)^2 \right)\\
&\simeq
-\frac{f_S(r) f_L}{\sigma_{T_L}^2}
+ \mathcal{O}\left( \left(\frac{f_L}{\sigma_{T_L}}\right)^2 \right).\\
\eal
\eeq

Next, we evaluate the foreground bias for the signed estimator. The stacked signed estimator including large- and small-scale foregrounds is
\beq
\hat{\tau}^{\rm Sgn}
=
-\frac{\sum \mathrm{sgn}(\delta T_L + f_L)(\delta T_S + f_S)}{\sum |\delta T_L + f_L|}.
\eeq
The bias to the stacked signed estimator to first order in $f_L/\sigma_{T_L}$ is
\beq
\label{eq:sgn_bias}
\bal
\hat{\tau}_{\rm bias}^{\rm sgn}(r)
&=
-\left<
\frac{\sum \mathrm{sgn}(\delta T_L + f_L)f_S(r)}
{\sum |\delta T_L|}  
\right>
+ \mathcal{O}\left( \left(\frac{f_L}{\sigma_{T_L}}\right)^2 \right)\\
&\simeq
-f_S(r)\frac{\left<\sum \mathrm{sgn}(\delta T_L + f_L)\right>}
{\left<\sum |\delta T_L|\right>}
+ \mathcal{O}\left( \left(\frac{f_L}{\sigma_{T_L}}\right)^2 \right).
\eal
\eeq

To determine the value of $\left<\sum\mathrm{sgn}(\delta T_L + f_L)\right>$, consider the case where $\delta T_L\sim\mathcal{N}(0, \sigma_{T_L})$ and $f_L$ is a small, positive, fixed value.
In this case, the distribution of $\delta T_L + f_L$ is just a translated Gaussian, i.e., $\delta T_L + f_L \sim\mathcal{N}(f_L, \sigma_{T_L})$.
Thus, the average value of the signed estimator is
\beq
\label{eq:sum_sgn}
\bal
\left<\sum\mathrm{sgn}(\delta T_L + f_L)\right>
&= A \int_{0}^{\infty} \mathrm{exp}\left[-\frac{(x-f_L)^2}{2\sigma_{T_L}^2}\right]dx\\
&\qquad - A\int_{-\infty}^0 \mathrm{exp}\left[-\frac{(x-f_L)^2}{2\sigma_{T_L}^2}\right]dx \\
&= A\int_{-f_L}^{f_L} \mathrm{exp}\left[-\frac{x^2}{2\sigma_{T_L}^2}\right]dx\\
&= \sqrt{\frac{2}{\pi}}\frac{f_L}{\sigma_{T_L}} + \mathcal{O}\left( \left(\frac{f_L}{\sigma_{T_L}}\right)^2 \right),
\eal
\eeq
where $A\equiv \frac{1}{\sqrt{2\pi}\sigma_{T_L}}$.

Substituting the result from Eq. \ref{eq:sum_sgn} into Eq. \ref{eq:sgn_bias} and using the relation $\left<|\mathcal{N}(\mu,\sigma)|\right>^2 = \frac{2}{\pi}\left<\mathcal{N}(\mu,\sigma)^2\right>$, we find
\beq
\bal
\hat{\tau}_{\rm bias}^{\rm Sgn}(r)
&\simeq
-\sqrt{\frac{2}{\pi}}\frac{f_L}{\sigma_{T_L}}\frac{f_S(r)}{\left<|\delta T_L|\right>}
+ \mathcal{O}\left( \left(\frac{f_L}{\sigma_{T_L}}\right)^2 \right)\\
&\simeq
-\frac{f_S(r) f_L}{\sigma_{T_L}^2}
+ \mathcal{O}\left( \left(\frac{f_L}{\sigma_{T_L}}\right)^2 \right).\\
\eal
\eeq
Thus, to first order in $f_L/\sigma_{T_L}$, the TI and signed estimators have the same sensitivity to foreground contamination.
For the signed estimator, this bias is only caused by regions where $|\delta T_L| < f_L$.
As sources are weighted by $\delta T_L$ in the TI estimator, these regions contribute very little to the stacked TI profile.
For the signed estimator, the mean bias can be completely mitigated by enforcing a minimum $\delta T_L$ cut such that $|\delta T_L| > |f_L|$ for all sources.

The analysis in this appendix assumes no overlapping of sources, which would contribute a non-trivial bias.
A detailed analysis of the effect of overlapping sources on the stacked estimators is left to future work.


\section{Simulations of correlated extragalactic foregrounds}
\label{app:fgsim}
\subsection{Simulating a lensing map correlated with $\tau$ map}
The deflection angle $\overrightarrow{\alpha}$ due to the gravitational lensing of photons is given by
\beq
\overrightarrow{\alpha} = \int \frac{d^2 \overrightarrow{\theta}}{\pi} \kappa (\overrightarrow{\theta}) \frac{\hat{\theta}}{\theta} \,
\eeq
where
\beq
\kappa(\overrightarrow{\theta}) = \frac{4\pi G \Sigma(\overrightarrow{\theta})}{c^2 a} \frac{d_L d_{LS}}{d_S}
\label{eq:sigmamaptokappamap}
\eeq
In the above equations, $\overrightarrow{\theta}$ are the 2D angles in the plane of the sky, $\kappa$ is the convergence field, $G$ is Newton's gravitational constant, $c$ is the speed of light, $a$ is the scale factor, $d_L$, $d_{LS}$, and $d_S$ are the comoving distance from observer to the lens, between the lens and the source, and to the source plane respectively. $\Sigma(\overrightarrow{\theta})$ is the comoving angular surface density. We therefore need $\Sigma(\overrightarrow{\theta})$ to get a lensing map $\kappa (\overrightarrow{\theta})$ correlated with the $\tau$ map. For this, we first assume a Gaussian projected profile for the angular surface density which we apply to the point-like galaxy maps. For CMASS galaxies with typical mass $M = 2 \times 10^{13} M_\odot$ at $z=0.55$, the virial radius is $\approx 1.6'$ \cite{Schaan2021a}. If we assume that the $3\sigma$ span of the Gaussian surface density profile should cover most of the profile and that all the mass falls within this then we have $\sigma = 1.6'/3$ i.e. angular surface density profile is a Gaussian with $\sigma \approx 0.53'$. In order to determine the normalization factor of this profile, we use the fact that $\int d^2 \overrightarrow{r} \Sigma(\overrightarrow{r}) = N_{\rm gal} M$ where $N_{\rm gal}$ is the total number of galaxies, $M$ is their mass, and $d^2 \overrightarrow{r} = {\rm pixel \: area (in \: steradians)} \times d_\chi^2$. Here the pixel area corresponds to the area of a single pixel of the map and $d_\chi$ is the average comoving distance to the galaxy sample. After getting this normalization, we have $\Sigma(\overrightarrow{\theta})$ which gives us $\kappa(\overrightarrow{\theta})$ using Eq.~\ref{eq:sigmamaptokappamap}.
Since the $\tau$ map is created using the same galaxy map as described in App.~\ref{app:tisims}, the $\kappa$ map and $\tau$ map are correlated.

\subsection{Simulating a tSZ map correlated with the $\tau$ map} \label{app:tszsims}
We make some simple assumptions to simulate a tSZ map correlated with the $\tau$ map. For example, we see from Fig.~8 of \cite{Schaan2021a} that the amplitude of the tSZ profile of CMASS galaxies is around $-10 \mu K \: {\rm arcmin}^2$ at 150 GHz. We then assume that the tSZ profile is Gaussian in nature centered at each galaxy location. The profile from Fig.~8 of \cite{Schaan2021a} extends out to $6'$ and is a beam convolved profile where the beam full width half maximum (FWHM) is $1.4'$. Therefore, debeamed FWHM for our simulated tSZ profiles is $\sqrt{6^2 - 1.4^2} \approx 5.8'$. Similar to lensing, in order to determine the correct normalization for the Gaussian profile, we use the fact that $\int d^2 \overrightarrow{\theta} T = N_{\rm gal} \times -10 \mu K \: {\rm arcmin}^2$ where $T$ is the tSZ profile in $\mu K$ and $\int d^2 \overrightarrow{\theta} T = ({\rm pixel \: area \: in \: {arcmin}^2}) \sum_i T_i$ where $\sum_i T_i$ is the temperature profile summed over all the pixels. The normalization is therefore determined such that this relation holds and we get our simulated tSZ map. 

We then calculate the power spectrum of our tSZ map and compare it with the expected power from galaxies with halo mass $5 \times 10^{12}M_\odot < M < 10^{14}M_\odot $ and redshift $0.4 < z < 0.7$, i.e. mass and redshift range similar to that of CMASS galaxies. We use the halo model predictions from \cite{Maniyar_2021} for this purpose. This prediction provides an upper limit on the expected power from the CMASS galaxies. As expected, we find that the tSZ power from our simulated map is below the halo model predictions and it also falls down much quicker on small scales. We call this map as `tSZ low'. In order to determine the upper limit for the tSZ bias to the $\tau$ QE, we thus simulate another tSZ map whose power spectrum roughly matches with the halo model predictions. For this, we simply convolve our galaxy map by a Gaussian beam with $\sigma = 0.88'$ instead of $\sigma=2.46'$ for our first tSZ map, such that the shape of its power spectrum roughly matches with that of halo model predictions. Then we multiply it by a factor of 1.3 on map level to match the amplitude of the power spectrum with predictions. Please note that these two numbers, namely the beam with $\sigma = 0.88'$ and extra normalization factor of 1.3 have been determined through trial and error to match the halo model prediction. This gives us a second tSZ map which we use to determine the highest expected bias from the tSZ emission coming from CMASS galaxies to cross-correlation of $\tau$ reconstructed with a QE with galaxies. We call this map as `tSZ high'.

\section{Simulation Methods for TI SNR Forecast}
\label{app:tisims}

\subsection{Signal maps}
\label{app:tisims_signal}

First, we create a $\tau$ map correlated with CMASS galaxy positions.
We choose map dimensions of $10^{\circ}\times 10^{\circ}$  as a balance between sufficient statistics and computational efficiency.
To make the map, we create a \textit{plate carr\'ee} (CAR) projection map of the $200^{\circ}< \text{RA} <210^{\circ}$, $10^{\circ}< \text{Dec} <20^{\circ}$ region with a pixel scale of 0.5'.
We choose this region to avoid any large masked regions in the CMASS sample footprint.
At each galaxy position, we draw a 2D isotropic Gaussian with FWHM=5' and amplitude of 0.000245.
These values are chosen based on the fit to the signed estimator stacked profile using ACT$\times$unWISE data \cite{Coulton_2024}.

As shown in Fig. \ref{fig:2_halo_profile}, stacking on the unWISE (and potentially  LSST) positions causes an excess signal from the two-halo term.
To avoid artificially boosting the signal for unWISE and LSST during stacking, we use the measured $\tau$ signal from the CMASS map for all forecasts.

Finally, we create a separate signal map for each of the ACT, SPT, SO and CMB-S4 forecasts by convolving the map with a Gaussian kernel with a FWHM of 1.6', 1.6', 1.4' and 1.0', respectively, to match the beam of the corresponding experiment's noise map (see below).

\subsection{Noise maps}
\label{app:tisims_noise}

The noise maps for the forecast consist of Gaussian random field realizations with power spectra corresponding to each CMB experiment: ACT, SPT, SO, and CMB-S4.
The spectra are plotted in Fig. \ref{fig:tt_power}.
For ACT, we use the total power spectrum measured on the \textit{Planck}+ACT needlet internal linear combination (ILC) map, which includes all the CMB anisotropies and is convolved with a 1.6' Gaussian beam \cite{Coulton2023_NILCmaps}.
For SPT, we use the 5-year projected ILC residual power spectrum curve including the noise and foreground contributions \cite{Raghunathan2023.954.83R, Sobrin_2022}.
For SO, we use forecast ILC power spectrum with the ``goal'' sensitivity specifications from \cite{SOScience}.
For CMB-S4, we use detector noise-only (i.e. no foregrounds) spectrum assuming $1\,\mu$K$\cdot$arcmin sensitivity.
We add an analytic lensed CMB contribution to the SPT, SO and CMB-S4 power spectra before convolving with a Gaussian beam of 1.6', 1.4' and 1.0', respectively.
The maps have the same dimensions, pixel scale and CAR projection as the signal maps.

\subsection{Filtering \& Stacking}

We give a basic overview of the filtering and stacking algorithm here but refer the reader to \cite{Schaan2021a} for further details.
To obtain unbiased $T_L$ and $T_S$ estimates, we apply scale-dependent filtering to our maps.
We find the compensated aperture photometry filter used in \cite{Schaan2021a} results in a significant noise bias. 
Instead, we apply low- and high-pass filtering to the maps to decorrelate the large- and small-scale temperature estimates:
\beq
    f^\text{low-pass}_\ell=\begin{cases}
			1, & \text{if $\ell<2000$}\\
            \cos\left(\frac{(\ell-2000)\pi}{300}\right) & \text{otherwise} \\
            0 &  \text{if $\ell>2150$}
		 \end{cases}
\label{eq:lpf}
\eeq
\beq
    f^\text{high-pass}_\ell=\begin{cases}
			0 & \text{if $\ell<2350$}\\
            \sin\left(\frac{(\ell-2350)\pi}{300}\right) & \text{otherwise} \\
            1 &  \text{if $\ell>2500$}.
		 \end{cases}
\label{eq:hpf}
\eeq

For the stacked analysis we need catalogs for each galaxy sample.
For the CMASS and unWISE samples, the catalog consists of the galaxies that lie in the region $200^{\circ}< \text{RA} <210^{\circ}$, $10^{\circ}< \text{Dec} <20^{\circ}$.
As the DC2 footprint does not cover this region, we simply take a 10 $\times$ 10 deg$^2$ patch from the DC2 footprint (specifically $55^{\circ}< \text{RA} <65^{\circ}$, $-40^{\circ}< \text{Dec} <-30^{\circ}$) and shift these positions to the above region.
Using real (or realistically simulated) galaxy positions ensures the effects of clustering are taken into account in our forecast.
The CMASS, unWISE, and DC2 samples have 7659, 515,095 and 11,904,577 galaxies in this region, respectively.

After applying the low- and high-pass filtering, we use bilinear interpolation to make small $\sim$10'$\times$10' cutouts of the temperature maps centered on each galaxy position.
We estimate $T_L$ for a galaxy position by measuring the mean temperature of the low-pass filtered map in a $\sim$9' disk centered on the galaxy.
The $T_S$ radial profile is estimated using nine annular ring aperture photometry filters centered on the galaxy with outer radii equally-spaced between 1' and 6'.
In each aperture $i$, we measure the mean temperature of the high-pass filtered map, $T_{S,i}$.
The estimated mean $\tau$ in each aperture is calculated as
\beq
\hat{\tau_i} = -\frac{T_L T_{S,i}}{(T_L)^2}.
\label{eq:onegal}
\eeq

The optimal stacked estimator is inverse-variance weighted, as shown in Sec. \ref{sec:ti_stack}.
In the forecast, we use uniform weighting for simplicity.
Thus, the stacked $\tau$ profile simplifies to
\beq
\hat{\tau} = -\frac{\sum_{j=1}^{N_{\rm gal}}T_{L,j} T_{S,j}}{\sum_{j=1}^{N_{\rm gal}}(T_{L,j})^2}.
\label{eq:taustack}
\eeq

\subsection{Covariance Estimation \& SNR Calculation}
\label{app:tisims_cov}

To estimate the covariance matrix, we simulate 128 independent noise map realizations, filter and stack in the same manner as above and measure a mean $\tau$ profile for each realization.
We calculate the covariance as the sample covariance over the 128 mean $\tau$ profiles.
The Hartlap correction \cite{Hartlap2007} to this covariance matrix is $(n_\mathrm{sims} - n_\mathrm{bins} - 2)/(n_\mathrm{sims} -1) = 1.08$ for 128 simulations and 9 profile bins.

Using simulations to estimate the covariance matrix is superior to bootstrap resampling as it accounts for the covariance added by the clustering of the galaxies.
We find the difference between the covariance estimated from simulations versus from 10,000 bootstrap samples is negligible for the CMASS sample.
However, bootstrap resampling underestimates the covariance by $\sim20$\% for the unWISE sample and by more than 50\% for the LSST sample, as shown in Fig. \ref{fig:snr_ratio}.

For a given experiment with noise covariance matrix, $C$, and mean $\tau$ profile, $\mathbf{v}$, we measure the detection SNR of the signal profile over the $10^{\circ}\times10^{\circ}$ map as
\beq
\left(\frac{S}{N}\right)^2_{\rm map} = \mathbf{v}^{\rm T} C^{-1} \mathbf{v}.
\eeq
We rescale this result to the full survey sky fraction and galaxy number density as shown in Table \ref{tab:ngal}.


\section{SNR calculation for the $\tau$ quadratic estimator}
\label{app:qesnr}

For the QE of $\tau$, the SNR for the cross-power spectrum of the $\tau$ map reconstructed using QE and galaxy map $g$ is calculated analytically as
\beq
\left( \frac{S}{N} \right)^2 = f_{\rm sky} \sum_{{\ell_b}_{\rm min}}^{{\ell_b}_{\rm max}}  \frac{(2{\ell_b}+1)  \Delta \ell \left(C_{\ell_b}^{\tau g}\right)^2} {\left(C_{\ell_b}^{\hat{\tau} g}\right)^2 + C_{\ell_b}^{\hat{\tau} \hat{\tau}} \times C_{\ell_b}^{g g}} \, ,
\label{eq:snr}
\eeq
where 
\beq
C_{\ell_b} = \frac{1}{\Delta \ell} \sum_{\ell \in [\ell_1, \ell_2]} C_\ell \, ,
\eeq
and $\Delta \ell$ is the bin width. Here, $C_{\ell_b}^{\tau g}$ is the cross-power spectrum between input $\tau$ map and $g$ map, $C_{\ell_b}^{\hat{\tau} \hat{\tau}} = C_{\ell_b}^{\tau \tau} + N_{\ell_b}^{\tau \tau}$ where $C_{\ell_b}^{\hat{\tau} \hat{\tau}}$ is the power spectrum of the input $\tau$ map and $N_{\ell_b}^{\tau \tau}$ is the reconstruction noise for QE given by Eq.~\ref{eq:varqe}. $C_{\ell_b}^{\hat{\tau} \hat{\tau}}$ is therefore the power spectrum of the reconstructed $\tau$ map. $C_{\ell_b}^{g g}$ is the total galaxy power spectrum including the clustering and shot noise components, and $f_{\rm sky}$ is the common sky fraction between the CMB map and galaxy survey.
We start with the $\tau$ map and CMASS galaxy map as described in App.~\ref{app:tisims}, and cross-correlate them to calculate the signal $C_\ell^{\tau g}$. For the noise part, we take into account different clustering and shot noise levels in the CMASS, unWISE, and LSST galaxy samples. For CMASS sample and LSST galaxies, we calculate the galaxy clustering power spectrum using a halo model approach where we make sure that our model matches with the clustering power spectrum of the CMASS maps used in cross-correlations. For unWISE galaxies, we use the measured power spectra for the `blue' sample \cite{Krolewski_2020,Krolewski_2021}.
The shot noise levels for each sample are calculated using the number density values provided in Tab.~\ref{tab:ngal}.

For different CMB experiments, the reconstruction noise for the quadratic estimator for $\tau$ is different and is calculated using Eq.~\ref{eq:varqe}. 

\clearpage 
\newcommand{\xg}{\bm{x}_{\rm g}}
\newcommand{\TL}{T^{\rm L}}
\newcommand{\TLl}{T^{\rm L}_{\ell}}
\newcommand{\TS}{T^{\rm S}}
\newcommand{\TSl}{T^{\rm S}_{\ell}}
\newcommand{\TSR}{T^{\rm SR}}
\newcommand{\dg}{\delta^{\rm g}}
\newcommand{\dgl}{\delta^{\rm g}_{\bm{\ell}_g}}
\newcommand{\vls}{\bm{L}}
\newcommand{\vll}{\bm{\ell}_{\rm L}}
\newcommand{\vlg}{\bm{\ell}_{\rm g}}
\newcommand{\tauhat}[1]{\ensuremath{\Hat{\tau}^{(#1)}}}
\newcommand{\ellscut}{\ensuremath{\ell^{\rm S}_{\rm cut}}}
\newcommand{\elllcut}{\ensuremath{\ell^{\rm L}_{\rm cut}}}

\newcommand\eqn[1]{Eq.~\ref{#1}}
\newcommand\fig[1]{Fig.~\ref{#1}}

\onecolumngrid
\thispagestyle{empty}
\begin{center}
    \textbf{\large{Erratum: A new ``temperature inversion'' estimator to detect CMB patchy screening by large-scale structure}}
\end{center}

\section*{The ``temperature inversion'' and ``signed'' screening estimators are biased in the presence of lensing}
\setcounter{equation}{0}
\renewcommand{\theequation}{I\arabic{equation}}
In Sec.~\ref{sec:ti_lens}, we claim the TI and signed estimators are unbiased in the presence of lensing.
In general this is not true; without adaptations, the estimators are only unbiased in the case of an infinitely small beam and careful filtering of the output reconstructed $\tau$ signal.
In this erratum we derive the source of the bias and present a minimal adaptation of the TI estimator to mitigate the lensing bias. 
The lensing bias to our anisotropic screening estimator was first identified in \cite{Hadzhiyska25, Sailer25}, along with more insights into the lensing bias, a more optimal lens-hardened screening estimator and associated updated detection signal-to-noise  (S/N) forecasts.

\section*{Deriving the lensing bias in harmonic space}
\label{sec:bias_ell}

As discussed in Sec.~\ref{sec:ti_lens}, the TI and signed estimators are unbiased by lensing in the absence of filtering.
However, as also shown in \cite{Ferraro16} for kSZ with projected fields, a subtle effect of filtering the temperature fields to form the estimator is that it allows the leakage of lensing into the measured $\tau$ signal.
In this section, we derive the source of this lensing bias, closely following \cite{Ferraro16}.
The main difference between the cases of the TI/signed estimators and kSZ with projected fields is that the TI/signed estimators have a different filter for each temperature leg.

We define the unlensed primary CMB temperature fluctuations as $T$ and the lensed temperature fluctuations as $\Tilde{T}$.
\begin{equation}
    \Tilde{T}^{\rm L/S}_{\vl} = F^{\rm L/S}_{\ell} b_{\ell} \Tilde{T}_{\vl} \equiv f^{\rm L/S}_{\ell} \Tilde{T}_{\vl},
\label{eq:tls}
\end{equation}
where $b_{\ell}$ is the beam and $F^{\rm L/S}_{\ell}$ is the low- (L) or high-pass (S) filter.
$F^{\rm L}_{\ell}$ is a tophat filter (or smooth variation thereof) defined by a maximum scale $\ell^{\rm L}_{\rm cut}$ and $F^{\rm S}_{\ell}$ by a minimum scale $\ell^{\rm S}_{\rm cut}$.
We define the product of the beam and each scale filter as $f^{\rm L/S}_{\ell}$.

\begin{figure}[h!]
    \centering
    \begin{tabular}{cc}
        \textbf{Without lensing mitigation} &
        \hspace{9pt} \textbf{With lensing mitigation}\\
        \includegraphics[width=0.48\linewidth]{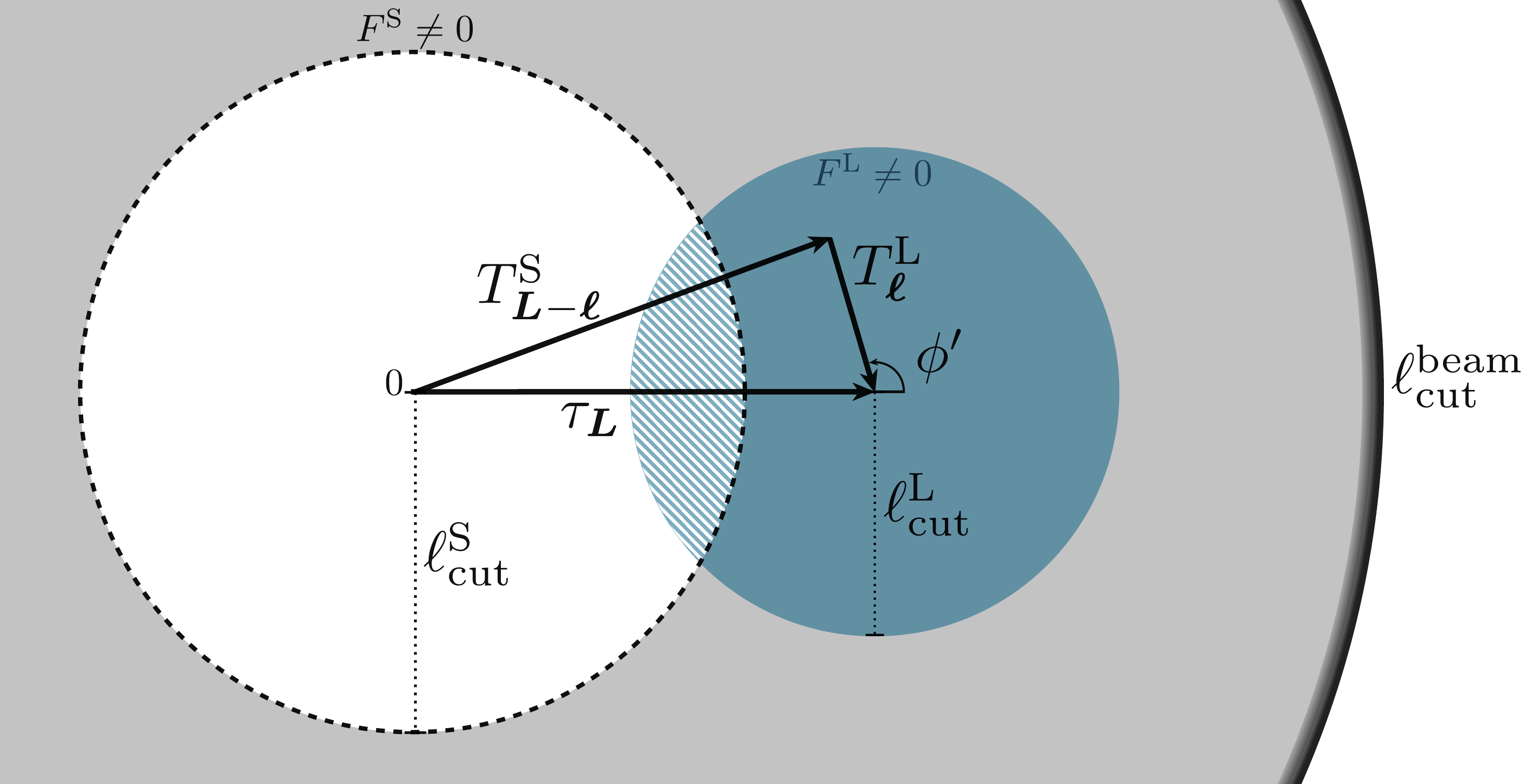} &
        \hspace{9pt}
        \includegraphics[width=0.48\linewidth]{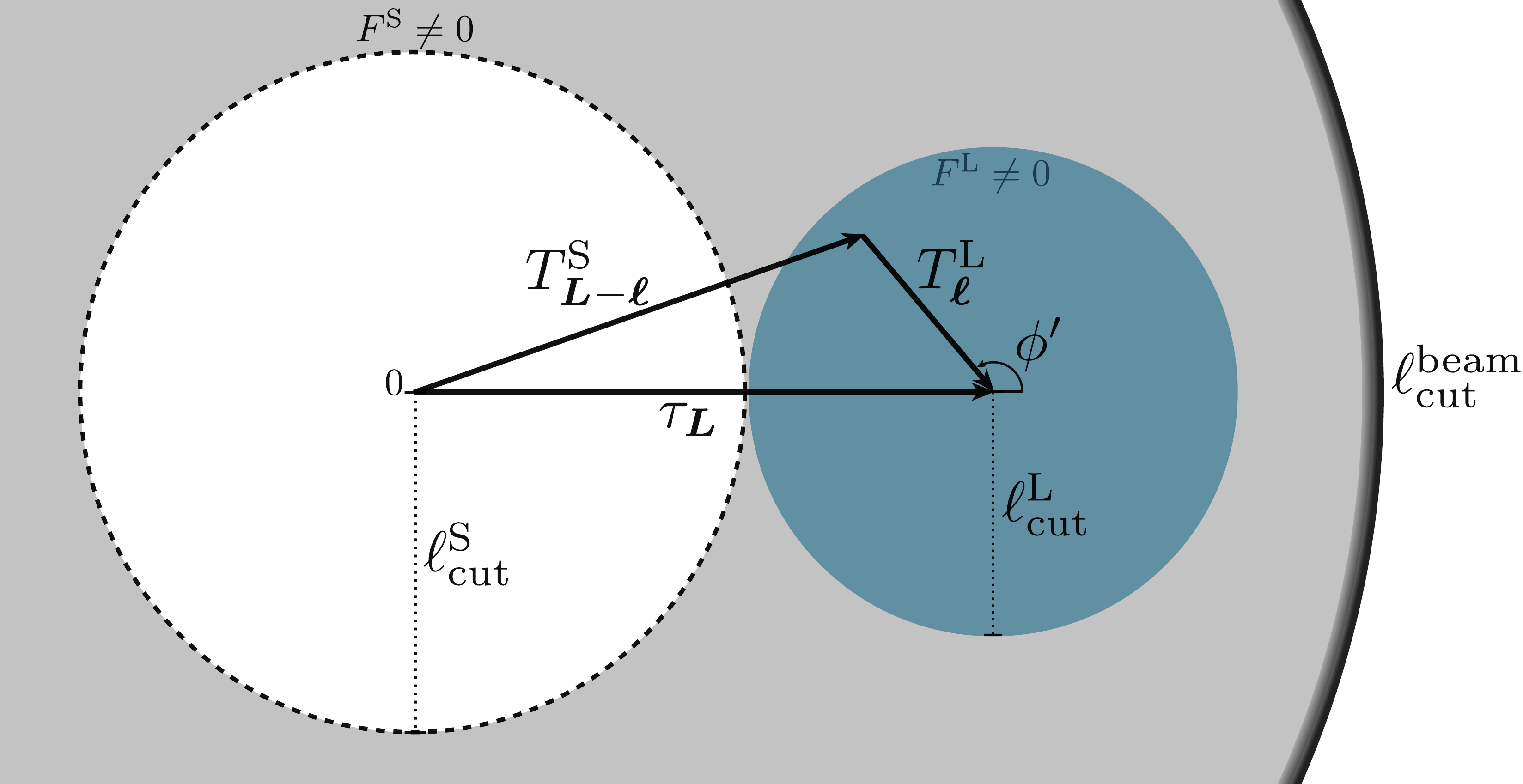}
    \end{tabular}
    \caption{
    Harmonic space diagram demonstrating the filtering scheme applied in the TI/signed estimators.
    The high-pass filter $F^{\rm S}$ with minimum scale \ellscut\  imposes $|\vL - \vl|>\ellscut$ (lighter gray shaded region), and the low-pass filter $F^{\rm L}$ with maximum scale \elllcut\ imposes $|\vl|<\elllcut$ (darker blue shaded region).
    The beam also imposes an upper limit of $\ell^{\rm beam}_{\rm cut}$ (rightmost solid black curve).
    In order for the $\phi$ integral in \eqn{eq:lensbias} to equal zero, the integration domain (i.e., where gray and blue overlap) must remain azimuthally symmetric and not cross the $F^{\rm S}$ or beam cutoff scales. Note $\phi'=\phi - \pi$ is used in the figures purely for visual clarity.\\
    \textbf{Left:} Without filtering $\tau_{\vL}$, the integration region can overlap with the modes filtered out by $F^{\rm S}$ (blue striped region), leading to a lensing bias.\\
    \textbf{Right:} The lensing bias can be mitigated by imposed by a filter on $\tau_\vL$ as proposed in \eqn{eq:sprime_filt}.
    }
    \label{fig:ell_fig}
\end{figure}

The stacked screening estimator at the positions of galaxies is of the form
\begin{align}
\begin{split}
    C_L^{\left(\Tilde{T}^{\rm L} \Tilde{T}^{\rm S}\right) \times \dg}
    & =  \int \frac{d^2\vl}{(2\pi)^2}  
    \left< \Tilde{T}^{\rm L}_{\vl} \Tilde{T}^{\rm S}_{\vL-\vl} 
    \dg_{- \bm L} \right>
     =  \int \frac{d^2\vl}{(2\pi)^2}  
     f^{\rm L}_{\ell} 
     f^{\rm S}_{|\vL-\vl|} 
     \left< \Tilde{T}_{\vl} \Tilde{T}_{\vls-\vl} \dg_{- \bm L} \right>.
    \label{eq:lsg_corr}
\end{split}
\end{align}
where $\dg$ is the galaxy density field.
The expectation value above may have a nonzero contribution due to lensing of the form
$\langle T T \phi \delta \rangle
=
\langle T T \rangle 
\langle \phi \delta \rangle
$,
resulting in a lensing bias.
More specifically, we Taylor expand the temperature fields to first order in the gradient of the lensing potential, $\nabla\varphi$:
\begin{equation}
    \Tilde{T}(\bm{x}) = T(\bm{x}) + \nabla\varphi(\bm{x}) \cdot \nabla T(\bm{x}) + \mathcal{O}((\nabla\varphi)^2).
\end{equation}
In Fourier space this is
\begin{equation}
    \Tilde{T}_{\vl} = T_{\vl} - \int \frac{d^2\vl'}{(2\pi)^2} \vl' \cdot (\vl-\vl') \varphi_{\vl'} T_{\vl - \vl'} + \mathcal{O}((\nabla\varphi)^2).
\end{equation}
We neglect the lensing contribution to $\TL$ but include it in $\TS$.
The lensing bias to \eqn{eq:lsg_corr} is then
\begin{align}
\begin{split}
    \int \frac{d^2\vl}{(2\pi)^2}  
    f^{\rm L}_{\ell} 
    f^{\rm S}_{| \vL-\vl |} 
    \left< [\nabla\varphi \cdot \nabla T(\bm{x})]_{\vL-\vl} T_{\vl} \dg_{-\bm L} \right>
    & = - \int \frac{d^2\vl}{(2\pi)^2}  
    f^{\rm L}_{\ell} 
    f^{\rm S}_{|\vL - \vl|}
    \int \frac{d^2\vl'}{(2\pi)^2} \vl' \cdot (\vL-\vl-\vl') \left< \varphi_{\vl'} T_{\vL-\vl-\vl'} T_{\vl} \dg_{-\bm L} \right>\\
    & = - \int \frac{d^2\vl}{(2\pi)^2}  
    f^{\rm L}_{\ell} 
    f^{\rm S}_{|\vL-\vl|}
    (\vL \cdot \vl) 
    C^{\varphi \dg}_{L} 
    C^{TT}_{\ell}
    \\
    & = -\frac{L C^{\varphi \dg}_{L}}{(2\pi)^2}
    \int d\ell \ell^2
    f^{\rm L}_{\ell}
    C^{TT}_{\ell}
    \int d\phi f^{\rm S}_{|\vL-\vl|} \cos{\phi},
\end{split}
\label{eq:lensbias}
\end{align}
where the second equality uses Wick's theorem and in the third equality $d^2\vl$ is expressed in polar coordinates with $\phi$ being the angle between $\vl$ and $\vL$.
\eqn{eq:lensbias} is zero by symmetry of the $\phi$ integral if the small-scale filter is constant where $f^{\rm L}$ is nonzero.
This is equivalent to the blue shaded region in Fig. \ref{fig:ell_fig} remaining a circle, i.e., crossing over neither the minimum high-pass filter scale $\ell^{\rm S}_{\rm cut}$ nor the effective beam cutoff scale $\ell^{\rm beam}_{\rm cut}$.
This can be enforced with a simple filter on the reconstructed $\vL$ leg
\begin{equation}
    S'_L = \left\{
        \begin{array}{ll}
         1 &\quad \ell^{\rm L}_{\rm cut} + \ell^{\rm S}_{\rm cut} < |\vL|<\ell^{\rm beam}_{\rm cut} - \ell^{\rm L}_{\rm cut},\\
         0 &\quad \mathrm{otherwise}.
        \end{array}
    \right.
    \label{eq:sprime_filt}
\end{equation}
Implementing the $S'$ filter mitigates lensing to first order at the cost of also filtering out some of the $\tau$ signal.

\section*{Real-space intuition on the lensing bias}

\begin{figure}
    \centering
    \includegraphics[width=0.7\linewidth]{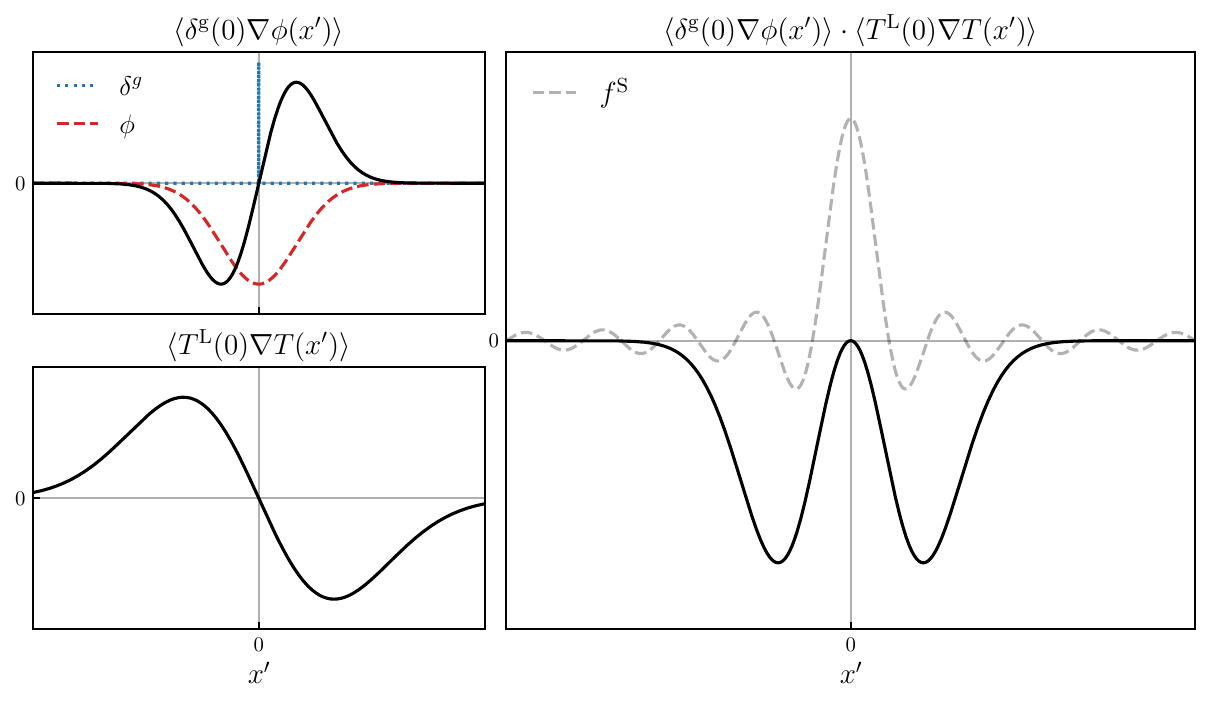}
    \caption{
    Schematic of the parity behavior of the quantities in \eqn{eq:taug} with respect to the separation $x'$.\\
    \textbf{Top left:} At the position of a galaxy at $x'=0$, $\dg$ (dotted blue) is a $\delta$ function.
    At the same location, the gravitational potential $\phi$ (dashed red) has a symmetric feature induced by the galaxy's halo.
    $\nabla \phi$ (solid black) is subsequently an odd function of $x'$. \\
    \textbf{Bottom left:} As shown in \eqn{eq:delphi_delt_parity}, for a Gaussian random field $\TL$, $\langle \TL(0) \nabla \TL(x')\rangle$ (solid black) is an odd function in $x'$ that is zero everywhere except within the correlation length ($\sim$degree scale) around $x'=0$. \\ 
    \textbf{Right:} Thus, the product $\langle \dg(0) \nabla \phi(\vx')  \rangle \cdot \langle T^{\rm L}(0) \nabla T(\vx') \rangle $ (solid black) is even.
    When convolved with the sinc-like filter $f^{\rm S}$ (dashed gray), $\langle \Hat{\tau}(0) \dg(0) \rangle$ will have nonzero response.
   }
    \label{fig:parity}
\end{figure}

To gain intuition on the source of the lensing bias, we also derive it here in real space.
The real-space analog to \eqn{eq:tls} is
\begin{align}
\begin{split}
    \Tilde{T}^{\rm L/S}(\vx) 
    &= \int d\vx' F^{\rm L/S}(\vx-\vx') \int d\vx'' b(\vx'-\vx'') \Tilde{T}(\vx'') \equiv \int d\vx' f^{\rm L/S}(\vx-\vx') \Tilde{T}(\vx').
\end{split}
\end{align}
As in the previous section, we consider only the lensing contribution to the small-scale $T^{\rm S}(\vx)$:
\begin{align}
\begin{split}
    T^{\rm S, lens}(\vx)
    & = (\nabla T \cdot \nabla \phi)^{\rm S}(\vx) = \int d\vx' f^{\rm S}(\vx - \vx') \nabla T(\vx') \cdot \nabla \phi(\vx').
\end{split}
\end{align}
Consider the lensing contribution to the zero-lag correlation $\langle \Hat{\tau}(0) \dg(0) \rangle$:
\begin{align}
    \langle \Hat{\tau}(0) \dg(0) \rangle
    & = \langle T^{\rm L}(0) (\nabla T \cdot \nabla \phi)^{\rm S}(0) \dg(0) \rangle.
\end{align}
In the absence of small-scale filtering we would get
\begin{align}
    \langle \Hat{\tau}(0) \dg(0) \rangle
    & = \underset{=0}{\underbrace{\langle T^{\rm L}(0) \nabla T(0) \rangle}} \cdot \underset{=0}{\underbrace{\langle \nabla \phi(0) \dg(0) \rangle}} = 0.
\end{align}
However, once we convolve with the small-scale filter $f^{\rm S}$, the nonzero-lag correlations must be considered.
It can be shown that, in general, $\langle T(0) \nabla T(\vx') \rangle$ and $\langle \dg(0) \nabla \phi(\vx') \rangle$ are odd functions of the separation $x'$:
\begin{equation}
    \langle T(0) \nabla T(\vx') \rangle = - \int \frac{d^2\vl}{(2\pi)^2} \vl C_{\ell}^{TT} \sin{\ell x'} ; \quad
    \langle \dg(0) \nabla \phi(\vx') \rangle = - \int \frac{d^2\vl}{(2\pi)^2} \vl C_{\ell}^{\dg\phi} \sin{\ell x'}.
    \label{eq:delphi_delt_parity}
\end{equation}
Therefore, with the small-scale filtering, we instead get
\begin{align}
\begin{split}
    \langle \Hat{\tau}(0) \dg(0) \rangle
    & = \int d\vx' \underset{\rm even}{\underbrace{f^{\rm S}(\vx')}} \underset{\rm odd}{\underbrace{\langle T^{\rm L}(0) \nabla T(\vx') \rangle}} \cdot \underset{\rm odd}{\underbrace{\langle \nabla \phi(\vx') \dg(0) \rangle}} \neq 0.
\end{split}
\label{eq:taug}
\end{align}
The parity behavior of each function in \eqn{eq:taug} as a function of $x'$ is shown in \fig{fig:parity}.
Essentially, convolution with the $f^{\rm S}$ filter averages the product of the two odd functions $\langle T^{\rm L}(0) \nabla T(\vx') \rangle$ and $\langle \nabla \phi(\vx') \dg(0) \rangle$ over $\sim$arcmin scales, resulting in a nonzero lensing bias.

\section*{Minimal method for mitigating lensing bias}

We want to understand how to effectively adapt our TI estimator to incorporate simple filtering that mitigates the present lensing bias.
The estimator proposed in the main text is of the form
\begin{equation}
    \tauhat{1}_{\rm (R)} = \frac{\sum_{\rm g} \TL(\xg) \TSR(\xg)}{\sum_{\rm g} (T^L(\xg))^2},
    \label{eq:tau1}
\end{equation}
where the summation is over galaxies and $\{\xg\}$ is the set of galaxy positions. $R$ denotes the application of a ring aperture photometry filter, such that $\tauhat{1}_{\rm (R)}$ is a single number for a given ring aperture inner/outer radius.
With the estimator in this form, it is possible but computationally inconvenient to apply an additional filter like \eqn{eq:sprime_filt} on the $\bm{L}$ modes for the reconstructed $\tau$.
However, we can implement a very similar alternative method.
Here we form a map $m(\bm{x})$ by multiplying the two filtered maps $\TL (\bm{x})$ and $\TS (\bm{x})$.
The $S'$ filter is applied to $m(\bm{x})$ to form $m^{S'}(\bm{x}) = (\TL (\bm{x}) \TS (\bm{x}))^{S'}$.
Before applying any ring aperture filters, stacking on galaxy positions results in a stacked reconstructed $\tau$ map
\begin{equation}
    \tauhat{2}(\bm{x}) = \frac{\sum_{\rm g} m^{S'}(\bm{x} + \xg)}{\sum_{\rm g} (\TL (\xg))^2}.
    \label{eq:tau3}
\end{equation}
We can then apply the ring aperture filter $R$ to this stacked map to form the estimator
\begin{align}
\begin{split}
    \tauhat{2}_{\rm (R)}
    & = \frac{\sum_{\rm g} m^{S'R}(\xg)}{\sum_{\rm g} (\TL (\xg))^2} 
    = \frac{\sum_{\rm g} [(\TL (\bm{x}) \TS (\bm{x}))^{S'}]^{R}(\xg)}{\sum_{\rm g} (\TL (\xg))^2}.
    \label{eq:tau4}
\end{split}
\end{align}
We note that this altered estimator does not reduce to \eqn{eq:tau1} when $S'$ is removed.
However, if $\TL$ does not vary on the scales of the $R$ aperture filter, then $\tauhat{2}_{\rm (R)} \simeq \tauhat{1}_{\rm (R)}$ when $S' = 1$ everywhere (and in any case, $\TL \simeq {\rm const.}$ for small scales is an assumption also underlying \tauhat{1}).

To confirm the normalization $\sum_{\rm g} (\TL (\xg))^2$ remains correct for \tauhat{2} and $\tauhat{2}_{\rm (R)}$, consider the numerator of \eqn{eq:tau3}:
\begin{align}
\begin{split}
        \left< (\TL \TS)^{S'} \right>
            & = \left< [\TL (\tau T^0)^{\rm S}]^{S'} \right>\\
            & = \left< W_L^{S'} \int \frac{d^2\bm{\ell}}{(2\pi)^2} \TL_{\bm{\ell}} \TS_{\bm{L}-\bm{\ell}}\right>\\
            & = \left< W_L^{S'} \int \frac{d^2\bm{\ell}}{(2\pi)^2} W^{\rm L}_{\ell}T^0_{\bm{\ell}} W^{\rm S}_{|\bm{L}-\bm{\ell}|}\int \frac{d^2\bm{\ell'}}{(2\pi)^2}\tau_{\bm{\ell'}}T^0_{\bm{L}-\bm{\ell}-\bm{\ell'}}\right>\\
            & =  W_L^{S'} \int \frac{d^2\bm{\ell}}{(2\pi)^2} W^{\rm L}_{\ell} W^{\rm S}_{|\bm{L}-\bm{\ell}|}\int \frac{d^2\bm{\ell'}}{(2\pi)^2}\tau_{\bm{\ell'}} C_{\ell}^{TT} (2\pi)^2 \delta^{\rm D}_{\bm{L}-\bm{\ell'}}\\
            & = \tau_{\bm{L}} W_L^{S'}\int \frac{d^2\bm{\ell}}{(2\pi)^2} W^{\rm L}_{\ell} W^{\rm S}_{|\bm{L}-\bm{\ell}|}C_{\ell}^{TT}.
    \label{eq:tt_norm}
\end{split}
\end{align}
The $W^{\rm L}_{\ell}$ filter imposes a upper limit to $|\bm{\ell}| < \ell^{\rm L}_{\rm cut}$, truncating the integral over harmonic space. In this limit, $|\bm{L}| \gg |\bm{\ell}|$, and thus the integral in the final line of \eqn{eq:tt_norm} becomes
\begin{align}
\begin{split}
        \int \frac{d^2\bm{\ell}}{(2\pi)^2} W^{\rm L}_{\ell} W^{\rm S}_{|\bm{L}-\bm{\ell}|}C_{\ell}^{TT}
        & = \int_{|\bm{\ell}| < \ell^{\rm L}_{\rm cut}} \frac{d^2\bm{\ell}}{(2\pi)^2} W^{\rm L}_{\ell} W^{\rm S}_{|\bm{L}-\bm{\ell}|}C_{\ell}^{TT}\\
        & \approx W^{\rm S}_L\int_{|\bm{\ell}| < \ell^{\rm L}_{\rm cut}} \frac{d^2\bm{\ell}}{(2\pi)^2} C_{\ell}^{TT} \\
        & \approx W^{\rm S}_L\left< (\TL)^2 \right>.
\end{split}
\end{align}
Thus, the normalization remains correct in the regime where $\tau$ is reconstructed only on small scales.

Ref.~\cite{Coulton_2024} v2 presents the updated  measurements using this filtering scheme for the signed screening estimator.
After lensing mitigation, the measurement significance is no longer sufficient for a detection.
As the screening S/N forecasts in the main text were scaled by the original measured signal, the forecasted S/N values have been overestimated as well.

\end{document}